\documentclass[
aps,
prd,
onecolumn,
nofootinbib,
superscriptaddress,
showkeys
]{revtex4-1}
\usepackage{comment}
\usepackage{amsmath}
\usepackage{amssymb}
\usepackage{bm}
\usepackage{mathrsfs}
\usepackage{graphicx}
\usepackage{epstopdf}
\usepackage{booktabs}
\usepackage{multirow}
\usepackage{float}
\usepackage{rotating}
\usepackage{cancel}
\usepackage{color}
\usepackage{comment}
\usepackage{enumitem}
\usepackage{subfigure}
\usepackage{hyperref}

\hypersetup{
	colorlinks=true,
	urlcolor=blue,
	linkcolor=blue,
	citecolor=blue
}

\newcommand{\doilink}[1]{%
  \href{https://doi.org/#1}{doi:#1}%
}
\newcommand{\bra}{\begin{array}}
\newcommand{\era}{\end{array}}

\newcommand{\beq}{\begin{equation}}
\newcommand{\eeq}{\end{equation}}

\newcommand{\beqar}{\begin{eqnarray}}
\newcommand{\eeqar}{\end{eqnarray}}

\newcommand{\be}{\begin{equation}}
\newcommand{\ee}{\end{equation}}

\newcommand{\bea}{\begin{eqnarray}}
\newcommand{\eea}{\end{eqnarray}}

\newcommand{\bd}{\begin{displaymath}}
\newcommand{\ed}{\end{displaymath}}

\newcommand{\h}{\newcommand{\h}{\newcommand{\h}{\hbar}}}

\begin{document}

\title{Optical and thermodynamic properties of Kerr-Bertotti-Robinson black holes}

\author{Hassan Hassanabadi}
\email{hha1349@gmail.com}
\affiliation{Physics Department, California State University, Fresno, CA 93740, USA}
\affiliation{Department of Physics, University of Hradec Kr\'alov\'e, Rokitansk\'eho 62, 500 03 Hradec Kr\'alov\'e, Czechia}
\affiliation{Department of Physics and Electronics, Khazar University, 41 Mahsati Str, 1096 Baku, Azerbaijan}

\author{Michael R. R. Good}
\email{muon@asu.edu}
\affiliation{Physics Department \& Energetic Cosmos Laboratory, Nazarbayev University, Astana 010000, Kazakhstan}
\affiliation{Leung Center for Cosmology \& Particle Astrophysics, National Taiwan University, Taipei 10617, Taiwan}
\affiliation{Beyond Center for Fundamental Concepts in Science, Arizona State University, Tempe, AZ 85287, USA}

\author{Soroush Zare}
\email{soroushzrg@gmail.com}
\affiliation{Helsinki Institute of Physics, University of Helsinki, P.O. Box 64, FI-00014, Helsinki, Finland}

\author{Orlando Luongo}
\email{orlando.luongo@unicam.it}
\affiliation{Universit\`a di Camerino, Via Madonna delle Carceri 9, 62032 Camerino, Italy}
\affiliation{Department of Nanoscale Science and Engineering, University at Albany-SUNY, Albany, New York 12222, USA}
\affiliation{INAF - Osservatorio Astronomico di Brera, Milano, Italy}
\affiliation{Al-Farabi Kazakh National University, Al-Farabi av. 71, 050040 Almaty, Kazakhstan}

\author{Fariba Kafikang}
\email{f.kafi19@yahoo.com}
\affiliation{Center for Theoretical Physics, Khazar University, 41 Mehseti Street, Baku, AZ-1096, Azerbaijan}

\begin{abstract}
\begin{comment}
We investigate the main thermodynamic and optical properties of rotating Kerr-Bertotti-Robinson black holes. To do so, we derive the horizon mass, Hawking temperature, fixed-$a$ Helmholtz-like free energy, heat capacity and, finally, its extremal remnant configurations. In all these cases, we show that the standard Kerr results are recovered when the Bertotti-Robinson parameter vanishes. In the weak-field regime, the leading corrections to the thermodynamic quantities appear at order $B^2$, whereas the extremal remnant mass remains unchanged at this order. Further, we discuss an effective thermodynamic interpretation in which the Bertotti-Robinson scale plays a role analogous to that of an Anti-de Sitter (AdS) phase, emphasizing an effective cosmological pressure. We analyze the ergosphere, photon motion, spherical photon orbits, and celestial shadow coordinates using the Hamilton-Jacobi formalism. Finally, we explore the magnetic shadow susceptibility, photon-region boundaries, ergosphere thickness and photon-ergosphere gap, characterizing how the external Bertotti-Robinson background modifies the black hole shadow, photon region, and rotational energy extraction zone.
\end{comment}
%\begin{abstract}
We investigate the thermodynamic and optical properties of Kerr--Bertotti--Robinson black holes, namely rotating black holes immersed in an external Bertotti--Robinson electromagnetic background. In the fixed-$a$ ensemble, we derive the horizon mass relation, the Hawking temperature, the entropy, the Helmholtz-type free energy, the heat capacity, and the extremal remnant configuration. These quantities reduce smoothly to their Kerr counterparts as $B\to0$. In the weak-field regime, the leading thermodynamic corrections arise at order $B^2$; the extremal radius is shifted at this order, whereas the remnant mass receives its first correction only at order $B^4$. We also introduce a formal AdS-like thermodynamic interpretation of the Bertotti--Robinson scale, treating the associated pressure as an effective response variable rather than a genuine cosmological pressure. Because the spacetime is not asymptotically flat, we further compute the finite-radius Komar mass and the Komar charge associated with the horizon generator. Using the Hamilton--Jacobi formalism, we derive the separated null-geodesic potentials, the impact parameters of spherical photon orbits, and the celestial coordinates of the shadow boundary for a finite-distance observer. We then characterize the photon-region boundaries, ergosphere thickness, photon--ergosphere gap, shadow area, and magnetic shadow susceptibility. Within the perturbative regime considered, the Bertotti--Robinson background decreases the averaged ergosphere thickness and shadow area, increases the photon--ergosphere gap, and produces a negative shadow susceptibility whose magnitude is enhanced by rotation.
\end{abstract}
%\end{abstract}

\maketitle

\tableofcontents

\section{Introduction}

Black holes provide a fundamental arena for testing gravitational physics in the strong-field regime \cite{strong-field-11,strong-field-12,strong-field-1,strong-field-2,strong-field-3,strong-field-4,strong-field-5,strong-field-6}. Among the exact solutions of general relativity, the Kerr geometry describes the exterior field of an uncharged rotating black hole and provides the standard framework for investigating horizons, ergospheres, photon motion, black-hole shadows, and thermodynamics \cite{8,9,10,11,12,13,14,15,16,17,18}. More broadly, the Kerr hypothesis posits that astrophysical black holes are described by either the Schwarzschild or Kerr spacetime, thereby underpinning the conventional modeling of compact objects in the Universe \cite{kerrhypothesis}.

Realistic black holes, however, are not expected to be completely isolated. They may be surrounded by electromagnetic fields \cite{19,20,21,22,23,24}, accreting matter \cite{orl5,orl1,orl3,28,29}, or external gravitational backgrounds \cite{30,31,32,33,34,35}. Observables associated with accretion disks can differ significantly when the underlying spacetime departs from the Kerr geometry \cite{orl1,orl2,orl3,orl4,orl5,orl6,orl7,orl8,orl9,orl10,orl11,orl12,orl13}. Similarly, studies of quasi-periodic oscillations predict substantial differences when the background spacetime is modified \cite{ORLANDO1,ORLANDO2,ORLANDO3,ORLANDO4}. These considerations motivate the systematic investigation of Kerr-like geometries embedded in nontrivial external fields, whose backreaction may alter both the thermodynamic properties and the optical response of the black hole.

Within this context, the Bertotti--Robinson spacetime is a classic exact solution of Einstein--Maxwell theory. It describes a homogeneous electromagnetic universe and is closely related to the product geometry $\mathrm{AdS}_2\times S^2$. It also arises in the near-horizon limit of extremally charged black holes. These properties have made Bertotti--Robinson backgrounds valuable in the study of electrovacuum geometries and exact solution-generating techniques.

More recently, Bertotti--Robinson electromagnetic embeddings have been shown to provide seeds for generating new vacuum spacetimes through Harrison-type transformations and magnetic inversion symmetries. In particular, starting from accelerating Bertotti--Robinson black holes, the external electromagnetic field can be removed while a nontrivial gravitational backreaction is retained in the metric. This construction demonstrates that Bertotti--Robinson backgrounds can leave observable signatures \cite{36,37,38,39,40,41}.

An important recent development was the construction of the Kerr--Bertotti--Robinson black hole as an exact stationary and axisymmetric solution of Einstein--Maxwell theory. This geometry describes a rotating Kerr black hole immersed in an external Bertotti--Robinson-type electromagnetic background and is characterized by the mass parameter $m$, the rotation parameter $a$, and the external-field parameter $B$. In the limit $B\to0$, the metric reduces to the standard Kerr geometry, whereas suitable limiting choices of the black-hole parameters connect the solution to the Bertotti--Robinson electromagnetic universe. Unlike an electrically charged Kerr--Newman black hole, this configuration does not require the black hole itself to carry an intrinsic electric charge. Nevertheless, the parameter $B$ modifies the geometry through both the external electromagnetic field and its gravitational backreaction \cite{new,42,43}.

Several properties of this recently introduced geometry have already been investigated \cite{44,45,46,47,48,49,50,51,52,53,54,55}. In particular, the separability of null geodesics, the structure of spherical photon orbits, and the resulting black-hole shadow have been analyzed in detail \cite{44,45,48,54}. These studies show that the Bertotti--Robinson parameter modifies the photon region and produces measurable deviations from the Kerr shadow \cite{44,51,54}. The optical appearance of the Kerr--Bertotti--Robinson black hole has also been examined under different illumination models, including celestial light sources and geometrically thin accretion disks. The results indicate that the spin and external-field parameters affect the observed image in distinct ways \cite{44,55}. Related analyses of magnetically driven synchrotron emission further show that the local electromagnetic environment can modify the disk-emission profile, lensing rings, and photon-ring substructure \cite{55}. Energy-extraction processes have likewise been considered, demonstrating that the external background alters both the ergoregion and the efficiency of rotational-energy extraction \cite{47}.

Despite this progress, several physical aspects of the Kerr--Bertotti--Robinson black hole remain unresolved. A particularly important issue is that the spacetime is not asymptotically flat. Consequently, conserved quantities such as the mass require a treatment different from that used in geometries approaching Minkowski spacetime at large distances. In particular, the Komar integral associated with the stationary Killing vector defines a finite-surface gravitational quantity that generally contains contributions from both the black hole and the external Bertotti--Robinson background.

In this work, we investigate the thermodynamic and optical properties of the Kerr--Bertotti--Robinson black hole within a unified framework. We first analyze the horizon structure and derive the horizon mass relation, the Hawking temperature, the entropy, the fixed-$a$ Helmholtz-type free energy, the heat capacity, and the extremal remnant configuration. We also discuss an effective thermodynamic interpretation in which the Bertotti--Robinson parameter defines an AdS-like scale. This interpretation is employed strictly as a formal diagnostic analogy and not as a genuine cosmological-pressure description. We then examine the Komar mass at finite radius and the Komar integral associated with the horizon-generating Killing vector.

We also investigate the optical sector. Using the Hamilton--Jacobi formalism, we derive the radial and angular photon potentials, the impact parameters associated with spherical photon orbits, and the celestial coordinates of the shadow boundary for a finite-distance observer. We then study the angular structure of the photon-region boundaries, the ergosurface, and the thickness of the ergosphere. Finally, we introduce diagnostic observables, including the photon--ergosphere gap, the shadow area, and the magnetic shadow susceptibility. In the weak-field regime, the leading corrections arise at order $B^2$, and all results reduce smoothly to their Kerr counterparts as $B\to0$. Together, these observables provide a systematic means of quantifying how an external Bertotti--Robinson electromagnetic background modifies rotating-black-hole thermodynamics, photon dynamics, and shadow morphology.

The paper is summarily organized as follows. In
Sec.~\ref{sec:geometry-horizon}, we introduce the
Kerr--Bertotti--Robinson geometry, establish the regular-axis
normalization, and analyze the horizon structure and horizon mass
relation. Section~\ref{sec:thermodynamics} develops the fixed-$a$
thermodynamic framework, including the Hawking temperature, entropy,
effective AdS-like response variables, Helmholtz-type free energy, and
heat capacity. The extremal configuration, remnant properties, and
admissible range of the Bertotti--Robinson parameter are discussed in
Sec.~\ref{sec:extremal-remnant}. In Sec.~\ref{sec:komar}, we evaluate
the finite-radius Komar mass and the Komar charge associated with the
horizon generator. Sections~\ref{sec:photon-motion} and
\ref{sec:photon-orbits} derive the null-geodesic equations, spherical
photon orbits, photon-region boundaries, and finite-distance celestial
coordinates. Section~\ref{sec:ergosphere} examines the ergosurface,
ergosphere thickness, and photon--ergosphere gap, while
Sec.~\ref{sec:shadow-observables} presents the shadow area and magnetic
shadow susceptibility. Finally, Sec.~\ref{sec:outlook} summarizes our
results and outlines future directions.

\section{Geometry and Horizon Structure}
\label{sec:geometry-horizon}
We consider the Kerr-Bertotti-Robinson spacetime described by the line element \cite{new}
\begin{equation}\label{eq-metric}
	ds^2=\frac{1}{\Omega^2}\left[
	-\frac{Q}{\Sigma}\left(dt-a\sin^2\theta\,d\phi\right)^2
	+\frac{\Sigma}{Q}\,dr^2
	+\frac{\Sigma}{P}\,d\theta^2
	+\frac{P}{\Sigma}\sin^2\theta
	\left(a\,dt-(r^2+a^2)d\phi\right)^2
	\right],
\end{equation}
where

\begin{subequations}
    \begin{align}
        &\Sigma=r^2+a^2\cos^2\theta,\\
        &P=1+B^2\left(m^2\frac{I_2}{I_1^2}-a^2\right)\cos^2\theta,\\
        &Q=(1+B^2r^2)\Delta(r),\label{eq-Q}\\
        &	\Omega^2=(1+B^2r^2)-B^2\Delta(r)\cos^2\theta,\\
        &	\Delta(r)=\left(1-B^2m^2\frac{I_2}{I_1^2}\right)r^2
	-2m\frac{I_2}{I_1}r+a^2.
    \end{align}
\end{subequations}
Here, $
	I_1=1-\frac{1}{2}B^2a^2$, $I_2=1-B^2a^2$ and it is useful to introduce
\begin{equation}\label{gamma}
	\gamma=\frac{I_2}{I_1}
	,\qquad \delta=\frac{I_2}{I_1^2}.
\end{equation}
To eliminate the possible conical defect on the symmetry axis, we introduce the regular azimuthal coordinate, and we consider the Kerr-Bertotti-Robinson spacetime in the regular-axis azimuthal normalization. The azimuthal coordinate is rescaled according to
$d\phi\longrightarrow \frac{d\varphi}{P_{0}}$, where \cite{Hu2026}
\begin{equation}
	P_0=P(0)=P(\pi)=1+B^2\left(m^2\delta-a^2\right).
	\label{eq:P0}
\end{equation}
and, for simplicity, the same symbol $\varphi$ is retained throughout.
Its period is fixed to $\Delta\varphi=2\pi$. 
Consequently, the covariant metric components transform as
$g_{t\varphi}
\longrightarrow
\frac{g_{t\varphi}}{P_{0}}$ and $g_{\varphi\varphi}
\longrightarrow\frac{g_{\varphi\varphi}}{P_{0}^{2}}$, whereas the corresponding inverse metric components transform as
$g^{t\varphi}
\longrightarrow
P_{0}\,g^{t\varphi}$ and $g^{\varphi\varphi} \longrightarrow P_{0}^{2} g^{\varphi\varphi}$.
The remaining components are unchanged:
$g_{tt}\longrightarrow g_{tt}$,
$g_{rr}\longrightarrow g_{rr}$,
$g_{\theta\theta}\longrightarrow g_{\theta\theta}$, and 
$g^{tt}\longrightarrow g^{tt}$.
Throughout this work, $m$ denotes the metric mass parameter, $a$ is the rotation parameter, and $B$ characterizes the external Bertotti-Robinson background. The outer horizon is denoted by $r_+$, and in the thermodynamic sections we also write $r_h\equiv r_+$. We define the dimensionless spin and magnetic parameters by
\begin{equation}
j\equiv \frac{a}{m},\qquad b\equiv Bm .
\end{equation}
Thus, weak-field corrections proportional to $B^2$ are equivalent to order $b^2$. The thermodynamic mass is denoted by $M$. In the effective extended description, $M$ is interpreted as enthalpy, $H=M$, whereas in the non-extended description, it is interpreted as internal energy. We use $\mathcal K$ for the Hamilton-Jacobi separation constant, reserving $K(j)$ for the complete elliptic integral of the first kind.

The event horizon is determined by the condition $Q(r_h)=0$. Since the prefactor $(1+B^2r_h^2)$ is strictly positive for real $B$ and $r_h$, the horizon condition reduces to $\Delta(r_h)=0$. Therefore, the horizon radius is 
\begin{equation}\label{eq:rhorizon}
	r_{\pm}
=
\frac{
m\gamma
\pm
\sqrt{
m^{2}\gamma^{2}
-
a^{2}\left(1-B^{2}m^{2}\delta\right)
}
}{
1-B^{2}m^{2}\delta
}.
\end{equation}
The outer event horizon is identified with the positive branch $r_h\equiv r_+$.

Equivalently, the same horizon equation can be inverted to express the black hole mass as a function of the horizon radius.
The outer horizon has the expansion
\begin{equation}
	r_{+}
	=
	r_{+}^{(0)}+B^{2}r_{+}^{(1)}+\mathcal{O}(B^{4}),
\end{equation}
with
\begin{equation}
	r_{+}^{(0)}
	=
	m+\sqrt{m^{2}-a^{2}},
	\qquad
	r_{+}^{(1)}
	=
	\frac{m}{2}
	\left(m+\sqrt{m^{2}-a^{2}}\right)^{2}.
\end{equation}
The physical positive-mass branch is therefore
\begin{equation}
	m_h(r_h)
=
\frac{
\sqrt{\gamma^{2}+B^{2}\delta\left(r_h^{2}+a^{2}\right)}
-\gamma
}{
B^{2}\delta\,r_h
}=\frac{
r_h^{2}+a^{2}
}{
r_h\left[
\gamma+
\sqrt{\gamma^{2}+B^{2}\delta\left(r_h^{2}+a^{2}\right)}
\right]
} .
\end{equation}
This expression gives the horizon mass as a function of the horizon radius.

In the Kerr limit $(B\to 0)$, one has $\gamma\to 1$, and in the above equation $m_h(r_h)$ goes to $
	\frac{r_h^2+a^2}{2r_h},
$
which is the standard Kerr horizon relation.

For a weak Bertotti-Robinson deformation, $Br_h\ll 1$ and $Ba\ll 1$, we expand
\begin{equation}
	\gamma=1-\frac{1}{2}B^2a^2+\mathcal{O}(B^4), \qquad \delta=1+O(B^4a^4).
\end{equation}
The horizon mass then becomes
\begin{equation}
	m_h(r_h)\simeq
	\frac{r_h^2+a^2}{2r_h}
	\left[
	1+\frac{B^2}{4}(a^2-r_h^2)
	\right]
	=
	\frac{r_h^2+a^2}{2r_h}
	+
	\frac{B^2(r_h^2+a^2)(a^2-r_h^2)}{8r_h}
	+\mathcal{O}(B^4).
\end{equation}
Thus, the Kerr horizon relation is recovered smoothly in the limit $B\to0$, while the leading correction induced by the Bertotti-Robinson background appears at order $B^2$.

\section{Thermodynamics}
\label{sec:thermodynamics}

Before analyzing the optical properties of the Kerr-Bertotti-Robinson geometry, we first discuss its thermodynamics. This section is devoted to the main thermodynamic quantities associated with the outer event horizon, with particular emphasis on the role played by the Bertotti-Robinson parameter $B$. We begin by deriving the Hawking temperature from the surface gravity and by showing how the standard Kerr result is recovered in the limit $B\to0$. We then introduce an effective AdS-like thermodynamic interpretation, in which the external Bertotti-Robinson scale is formally associated with a pressure-like variable. 

Since the spacetime is not asymptotically flat and the conserved charges require a separate analysis, we fix $a$ and distinguish this effective description from the standard fixed-$J$ thermodynamics of rotating black holes. Within this framework, we define the corresponding response volume, magnetic response coefficient, Helmholtz-like free energy, and heat capacity. 

These quantities provide the basic thermodynamic requirements for characterizing the stability properties and remnant structure of the black hole.

\subsection{Hawking Temperature}

The Hawking temperature is determined by the surface gravity at the outer event horizon \cite{56,57}. For the Kerr-Bertotti-Robinson geometry, the surface gravity can be written in terms of the radial metric function $Q(r)$ as

\begin{equation}
	\kappa_h=\frac{Q'(r_h)}{2(r_h^2+a^2)} .
\end{equation}
Because the Kerr-Bertotti-Robinson spacetime is not asymptotically	flat, the normalization of the stationary Killing vector cannot be fixed by imposing $\xi^\mu \xi_\mu\to -1$ at spatial infinity. In this work we use the normalization inherited from the original form of the
	Kerr-Bertotti-Robinson metric, namely the Killing field
	$\xi=\partial_t$ associated with the Kerr-like time coordinate $t$.
	With this convention, the horizon generator is $\chi=\partial_t+\Omega_h\partial_\phi$, and the surface gravity is
	\begin{equation}
		\kappa_h^{(t)}=\frac{Q'(r_h)}{2(r_h^2+a^2)} .
		\label{eq:kappa}
	\end{equation}
	Therefore the Hawking temperature quoted below,
	\begin{equation}
		T_h^{(t)}=\frac{\kappa_h^{(t)}}{2\pi},
		\label{eq:Th}
	\end{equation}
	should be understood as the temperature measured with respect to this
	chosen time normalization. For simplicity, we denote $T_h^{(t)}\equiv T_h$ in the following. A different constant rescaling of the time
	coordinate would rescale $\kappa_h$ and $T_h$ by the same constant.
	The present convention is chosen because it reduces smoothly to the
	standard Kerr normalization in the limit $B\to0$. The Hawking temperature is therefore

\begin{equation}
	T_h=\frac{\kappa_h}{2\pi}
	=\frac{Q'(r_h)}{4\pi(r_h^2+a^2)} .
\end{equation}
Using $Q'(r_h) = (1 + B^2r_h^2)\Delta'(r_h),$ which follows from the horizon condition $\Delta(r_h) = 0$ has been used \cite{58,59}. Hence,
\begin{equation}\label{eq:temprature}
	T_h =\frac{1 + B^2r_h^2}{4\pi(r_h^2 + a^2)}
	\left[2\left(1 - B^2m^2\delta\right)r_h - 2m\gamma\right]=\frac{(1+B^2r_h^2)(\gamma r_h^2-a^2\zeta_h)}
	{2\pi r_h(r_h^2+a^2)(\gamma+\zeta_h)} ,
\end{equation}
where

\begin{equation}\label{eq-hh}
	\zeta_h=\sqrt{\gamma^2+B^2\delta(r_h^2+a^2)} .
\end{equation}

In the limit $B\to 0$, one finds $\zeta_h\to 1$, yielding $T_h(r_h)$ reduces to $\frac{r_h^2-a^2}
{4\pi r_h(r_h^2+a^2)} ,$ which is precisely the Hawking temperature of the Kerr black hole.

For weak Bertotti-Robinson deformation, namely $Br_h\ll1$ and $Ba\ll1$, we expand the exact expression up to order $B^2$. This gives

\begin{equation}
	T_h =
	\frac{r_h^2-a^2}
	{4\pi r_h(r_h^2+a^2)}
	+
	\frac{B^2\left(3r_h^4-6a^2r_h^2-a^4\right)}
	{16\pi r_h(r_h^2+a^2)}
	+\mathcal{O}(B^4).
\end{equation}

\begin{comment}
Equivalently,

\begin{equation}
	T_h\simeq T_h^{\rm Kerr}+B^2\delta T_h ,
\end{equation}

where

\begin{equation}
	\delta T_h=
	\frac{3r_h^4-6a^2r_h^2-a^4}
	{16\pi r_h(r_h^2+a^2)} .
\end{equation}
\end{comment}

Thus, the leading correction to the Hawking temperature induced by the Bertotti-Robinson background appears at order $B^2$, while the standard Kerr result is smoothly recovered in the limit $B\to0$.

\subsection{Effective AdS-like pressure and effective response volume}\label{Sec:EffectiveAdS-likePressureandThermodynamicVolume}

For perturbative thermodynamic calculations, we restrict ourselves to the weak-field regime
\begin{equation}
	B^{2}a^{2}\ll 1,\qquad B^{2}m^{2}\ll 1.
\end{equation}

In this limit, the metric functions contain factors such as $1 + B^{2}r^{2}$, which resemble the AdS factor $1 + r^{2}/\ell^{2}$. However, this analogy is purely formal: the parameter $B$ characterizes the external Bertotti-Robinson electromagnetic background, not a fundamental cosmological constant. Nevertheless, one may define an effective length scale $\ell_{B} = 1/B$, an effective cosmological constant $\Lambda_{\mathrm{eff}} = -3B^{2}$, and an effective pressure
\begin{equation}
	\mathcal{P}_B = -\frac{\Lambda_{\mathrm{eff}}}{8\pi} = \frac{3B^2}{8\pi},
\end{equation}
which has the correct dimension of effective pressure variable in geometrized units ($[\mathcal{P}_B] = L^{-2}$). The quantity $\mathcal{P}_B$ is introduced only as an effective bookkeeping variable that parametrizes the response of the horizon
thermodynamics to variations of the Bertotti-Robinson parameter $B$.
It should not be interpreted as a genuine cosmological pressure. Indeed,
the Bertotti-Robinson background is not a four-dimensional Anti--de
Sitter spacetime; rather, in the appropriate limit it approaches the
product geometry $\mathrm{AdS}_2 \times S^2$. Consequently,
$\mathcal{P}_B$ is employed solely as a convenient response parameter
for describing the $B^2$-dependence of the thermodynamic quantities,
without implying the existence of a physical cosmological constant or an
extended black-hole thermodynamic phase space of the Kerr--AdS type.This construction provides a useful way to quantify the response of the black hole to the external Bertotti-Robinson scale, but it should not be interpreted as evidence that the spacetime has a true cosmological pressure of Kerr-AdS type.

In the non-extended thermodynamic description, the black hole mass is interpreted as the internal energy, $U = M$. If, instead, the Bertotti-Robinson parameter is promoted to an effective pressure variable, then the mass is naturally interpreted as enthalpy, $H = M$, and the internal energy is
\begin{equation}
	U = H - \mathcal{P}_B V_B.
\end{equation}

The effective response volume conjugate to $\mathcal{P}_B$ is defined by
\begin{equation}
	V_{B} ^{(a)}= \left(\frac{\partial M}{\partial\mathcal{P}_{B}}\right)_{S,a} = \frac{8\pi}{3}\left(\frac{\partial M}{\partial B^{2}}\right)_{S,a}.
\end{equation}

Therefore,
\begin{equation}\label{eq:VB}
	V_{B}^{(a)}
\longrightarrow
\frac{8\pi}{3}
\left[
\left(
\frac{\partial M}{\partial B^{2}}
\right)_{r_{h},a}
-
\left(
\frac{\partial M}{\partial r_{h}}
\right)_{B,a}
\frac{
\left(
\frac{\partial S_{h}}{\partial B^{2}}
\right)_{r_{h},a}
}{
\left(
\frac{\partial S_{h}}{\partial r_{h}}
\right)_{B,a}
}
\right].
\end{equation}
Thus, $V_B^{(a)}$ measures how the mass changes when the Bertotti-Robinson parameter is varied while the entropy $S$ and the spin parameter $a$ are held fixed. It should therefore be regarded as an effective fixed-spin response volume, rather than as the usual effective response volume defined at fixed angular momentum $J$,
\begin{equation}
V_B^{(J)}=\left(\frac{\partial M}{\partial \mathcal{P}_B}\right)_{S,J},
\end{equation}
which is different from $V_B^{(a)}$. In the present work we do not compute $V^{(J)}_B$. The quantity
	$V^{(a)}_B$ is used only as an effective fixed-spin response to the
	external Bertotti-Robinson scale, not as the standard thermodynamic
	volume of a fixed-$J$ extended first law.

\subsection{Magnetic response in the fixed-\texorpdfstring{$a$}{a} ensemble}
	
	The magnetic response associated with the fixed-$a$ ensemble is
	defined by
\begin{equation}
	\Psi_B^{(a)}
	=
	\left(\frac{\partial M}{\partial B}\right)_{S_h,a}.
\end{equation}
	Equivalently, using the effective pressure variable
	$P_B=3B^2/(8\pi)$, it can be written as
\begin{equation}
	\Psi_B^{(a)}
	=
	\frac{3B}{4\pi}V_B^{(a)} .
\end{equation}
	This quantity plays the role of a magnetization-like response to the
	external Bertotti-Robinson background. It should not be confused with
	the fixed-$J$ conjugate
\begin{equation}
	\Psi_B^{(J)}
	=
	\left(\frac{\partial M}{\partial B}\right)_{S_h,J},
\end{equation}
	which would belong to a different thermodynamic ensemble.
\subsection{Thermodynamic potentials in the fixed-\texorpdfstring{$a$}{a} ensemble}

We now summarize the thermodynamic quantities that will be used in the
fixed-$a$ description adopted in this work. At the horizon,
$r=r_h$, one has $Q(r_h)=0$. The horizon generator is written as $\chi=\partial_t+\Omega_h\partial_\phi$, where the angular velocity of the horizon is obtained from the usual
geometrical condition that $\chi^\mu\chi_\mu=0$ on the horizon. This gives
\begin{equation}
	\Omega_h=\frac{P_{0}a}{r_h^2+a^2}.
	\label{eq:Omega_h}
\end{equation}
This expression has the same functional form as in the Kerr geometry \cite{58},
although the horizon radius $r_h$ is modified by the
Bertotti-Robinson parameter. We emphasize, however, that in the present
non-asymptotically flat spacetime $\Omega_h$ should be understood as the geometrical angular velocity of the horizon. It should not be
automatically identified with the thermodynamic conjugate
$(\partial M/\partial J)_{S,B}$ unless a consistent thermodynamic
potential $M(S,J,B)$ is independently constructed.

The horizon entropy is obtained from the Bekenstein-Hawking area law \cite{56}, $S_h=\frac{A_h}{4}$.
Using the metric induced on the horizon and taking the azimuthal
coordinate to have the standard period $\Delta\phi=2\pi$, the horizon
area is
\begin{equation}
	A_h=\frac{4\pi(r_h^2+a^2)}{P_{0}(1+B^2r_h^2)}.
	\label{eq:Ah}
\end{equation}
Thus, the entropy reads
\begin{equation}
	S_h=\frac{\pi(r_h^2+a^2)}{P_{0}(1+B^2r_h^2)}.
	\label{eq:entropy}
\end{equation}
Easily, for $B\to0$, one recovers the Kerr case, i.e., 
$S_h=\pi(r_h^2+a^2)$.

The horizon relation can be inverted to define the effective horizon
mass
\begin{equation}
	M(r_h,a,B)\equiv m_h(r_h)
	=\frac{r_h^2+a^2}{ r_h\left(\gamma+\zeta_h\right)}.
	\label{eq:M_h}
\end{equation}

This quantity is the mass parameter determined by the horizon equation.
Since the Kerr-Bertotti-Robinson spacetime is not asymptotically flat
in the ordinary Kerr sense, this horizon mass should not be identified
automatically with an ADM mass. A complete construction of the physical
mass and angular momentum would require a separate conserved charge
analysis, possibly including an appropriate background subtraction.

For this reason, in the present work, we do not formulate a standard
fixed-$J$ first law of the form $dM=T_h\,dS_h+\Omega_h\,dJ+\Psi_B\,dB$. Such a relation would require an independently derived thermodynamic
function $M(S,J,B)$, from which
\begin{align}
	T&=\left(\frac{\partial M}{\partial S}\right)_{J,B},\qquad
	\Omega=\left(\frac{\partial M}{\partial J}\right)_{S,B},\qquad
	\Psi_B^{(J)}
	=
	\left(\frac{\partial M}{\partial B}\right)_{S,J}
	\label{eq:derivatives_J}
\end{align}
could be obtained consistently. Instead, we work in the fixed-$a$
ensemble. The response of the horizon mass to the external
Bertotti-Robinson parameter is then characterized by the fixed-$a$
magnetic response coefficient
\begin{equation}
	\Psi_B^{(a)}
	=
	\left(\frac{\partial M}{\partial B}\right)_{S_h,a}.
	\label{eq:Psi_a}
\end{equation}
Since $J$ has not been independently derived, this fixed-$a$ response is not interpreted as a fixed-$J$ thermodynamic conjugate.

Similarly, we do not use the Kerr relation $J=aM$ as a fundamental
definition of the angular momentum. When this relation is mentioned, it
should be understood only as a Kerr-limit motivated effective relation,
valid as a diagnostic analogy rather than as a first-principles
conserved charge for the non-asymptotically flat geometry.

For comparison, if one wished to work in a rotating grand-canonical
ensemble, the corresponding potential would be
\begin{equation}
\mathcal G=M-T_hS_h-\Omega_h J .
\end{equation}
This is the quantity usually called the rotating Gibbs free energy, or
grand-canonical free energy, for a rotating black hole. If one formally inserts the Kerr-like effective relation $J=aM$, one obtains
\begin{equation}
\mathcal G_{\rm eff}=F_a-\Omega_h aM
=
\frac{2P_{0}(r_h^2+a^2)+
	a^2\zeta_h-\gamma r_h^2-2P_{0}^2 a^2
}{
	2 P_{0}r_h(\gamma+\zeta_h)
}.
\end{equation}
However, because $J$ has not been derived independently as a conserved
charge for the Kerr-Bertotti-Robinson spacetime, we do not use
$\mathcal G_{\rm eff}$ as the fundamental thermodynamic potential in
this work. The quantity used in the following analysis is the
fixed-$a$ Helmholtz-type free energy $F_a$. Substituting the exact expressions for
$M$, $T_h$, and $S_h$, one obtains
\begin{equation}
	F_a(r_h)
	=
	\frac{
		2P_{0}(r_h^2+a^2)+a^2\zeta_h-\gamma r_h^2
	}{
		2 P_{0}r_h(\gamma+\zeta_h)
	}.
	\label{eq:Fa}
\end{equation}
In the Kerr limit $B\to0$, where $\gamma\to1$ and
$\zeta_h\to1$, this reduces to
\begin{equation}
	F_a^{\rm Kerr}(r_h)
	=
	\frac{r_h^2+3a^2}{4r_h}.
	\label{eq:Fa_Kerr}
\end{equation}
Thus, the Bertotti-Robinson background modifies the Helmholtz-type
free energy through its effect on the horizon mass, Hawking temperature,
and entropy.

For weak Bertotti-Robinson deformation, $Br_h\ll1$ and $Ba\ll1$,
the fixed-$a$ Helmholtz-type free energy expands as
\begin{comment}
\begin{equation}
	F_a(r_h)	=	\frac{r_h^2+3a^2}{4r_h}
	+\frac{B^2}{16r_h}
	\left(	-r_h^4+2a^2r_h^2+3a^4
	\right)	+O(B^4).
	\label{eq:Fa_expand}
\end{equation}
\end{comment}
\begin{equation}
F_a(r_h)
=
\frac{r_h^2+3a^2}{4r_h}
-
\frac{B^2a^2\left(r_h^4-6a^2r_h^2+a^4\right)}
{16r_h^3}
+\mathcal{O}(B^4).
\label{eq:Fa_expand}
\end{equation}
%The $B^2$ term gives the leading Bertotti-Robinson correction.

In obtaining this expansion, the $B^2$ dependence of the
regular-axis normalization has been retained. Indeed, using the
zeroth-order horizon relation
$M_0=(r_h^2+a^2)/(2r_h)$, one finds
\[
P_0
=
1+\frac{B^2(r_h^2-a^2)^2}{4r_h^2}
+\mathcal{O}(B^4).
\]
Consequently, the order-$B^2$ correction to $F_a$ vanishes in the
nonrotating limit $a\to0$.
\subsection{Heat capacity}
    
The heat capacity is one of the most useful thermodynamic quantities for studying the local stability of black holes. It measures how the entropy, or equivalently the horizon area, changes with temperature under a given thermodynamic ensemble. A positive heat capacity indicates that the black hole can remain in local thermal equilibrium with its environment, whereas a negative heat capacity signals local thermodynamic instability. Moreover, divergences of the heat capacity usually mark possible second-order phase transitions or Davies-type critical points \cite{58,59,60}.

\begin{equation}
	C_{a,B} = T_{h}\frac{(\partial S_{h} / \partial r_{h})_{a,B}}{(\partial T_{h} / \partial r_{h})_{a,B}}.
\end{equation}

Therefore, we have

\begin{equation}
	C_{a,B} = \frac{dS_{h} / dr_{h}}{D_T(r_{h})}, \qquad \frac{dS_{h}}{dr_{h}}
=
\frac{
2\pi r_{h}\left(1-B^{2}a^{2}\right)
}{
P_{0}\left(1+B^{2}r_{h}^{2}\right)^{2}
}
-
\frac{
2\pi B^{2}\delta\,M
\left(
\frac{\partial M}{\partial r_{h}}
\right)_{B,a}
\left(r_{h}^{2}+a^{2}\right)
}{
P_{0}^{2}\left(1+B^{2}r_{h}^{2}\right)
}.
\end{equation}

where $D_T(r_{h}) = \frac{d\ln T_{h}}{dr_{h}}$ gives

\begin{equation}
D_T(r_{h})	 = \frac{2B^{2}r_{h}}{1 + B^{2}r_{h}^{2}} + \frac{2\gamma r_{h} - \frac{a^{2}B^{2}\delta r_{h}}{\zeta_{h}}}{\gamma r_{h}^{2} - a^{2}\zeta_{h}} - \frac{1}{r_{h}} - \frac{2r_{h}}{r_{h}^{2} + a^{2}} - \frac{B^{2}\delta r_{h}}{\zeta_{h}(\gamma + \zeta_{h})}.
\end{equation}

The zeros and divergences of $C_{a,B}$ contain important physical information. The heat capacity vanishes when the black hole approaches a zero-temperature configuration, corresponding to the extremal remnant discussed in the next section. On the other hand, the heat capacity diverges when $D_T(r_{h}) = 0. $

Such divergences signal possible second-order phase transitions or changes in the black hole's local thermodynamic stability.

The black hole is locally thermodynamically stable for $C_{a,B} > 0$ and unstable for $C_{a,B} < 0$.

Therefore, the Bertotti-Robinson parameter affects the stability structure through both the entropy factor and the temperature derivative encoded in $D_T(r_{h})$.

\section{Extremal Configuration and Remnant}
\label{sec:extremal-remnant}
The extremal configuration describes the limiting state in which the black hole reaches zero Hawking temperature and the evaporation process terminates in a remnant. In this regime, the horizon becomes degenerate, and the thermodynamic temperature vanishes at the event horizon. To determine this configuration, we use the Hawking temperature obtained in Eq. \eqref{eq:temprature} and impose the condition $T_h(r_h)=0$ \cite{56,57}.

Therefore,
\begin{equation}
\gamma r^2_h=a^2\zeta_{h} .
\end{equation}
which leads to the remnant horizon radius
\begin{equation}
	r_{\rm rem}=\frac{a}{\sqrt{1-a^2B^2}}.
\end{equation}

Therefore, the Bertotti-Robinson background shifts the extremal radius away from its Kerr value.

\begin{equation}\label{eq:mrem}
	m_{\rm rem}=
	\frac{
		a\left(1-\frac{1}{2}B^2a^2\right)
	}{
		\sqrt{1-B^2a^2}
	}.
\end{equation}

In the Kerr limit ($B\to0$), the remnant radius and mass reduce to $	r_{\rm rem}\to a,$ and $	m_{\rm rem}\to a,$ which recovers the standard extremal Kerr relation.

For weak Bertotti-Robinson deformation ($Ba\ll1$), the remnant radius expands as

\begin{equation}
	r_{\rm rem}
	\simeq
	a\left(1+\frac{1}{2}B^2a^2\right)
	+\mathcal{O}(B^4).
\end{equation}

The remnant mass expands as

\begin{equation}
	m_{\rm rem}
	\simeq
	a\left[
	1+\frac{1}{8}B^4a^4
	+\mathcal{O}(B^6a^6)
	\right].
\end{equation}

Hence, the remnant radius receives its leading correction at order $B^2$, while the remnant mass remains unchanged at this order and acquires its first correction only at order $B^4$.

\subsection{Constraint on the magnetic field}

\begin{comment}
The physical range of the Bertotti-Robinson parameter can be constrained by requiring that the thermodynamic and geometrical quantities remain regular. Positivity and regularity of the entropy additionally require
$P_{0}>0$. In the weak-field regime, this condition is automatically satisfied. Outside the perturbative regime, however, it must be checked together
with the regularity condition
$1-B^{2}a^{2}>0$.
Therefore, the physically admissible parameter region must satisfy
$P_{0}>0$, and  $1-B^{2}a^{2}>0$.

Therefore, the positivity of entropy alone does not restrict the allowed values of $B$.

To avoid the divergence in Eq. \eqref{eq:mrem} and preserve the physical positive-mass branch, one requires $1-B^2 a^2>0$.

This gives the bound $	B^2a^2<1,$ or equivalently, $	|B|<\frac{1}{|a|}.$ Thus, the physically admissible parameter region is $	|Ba|<1 .$

This condition guarantees that the relevant metric functions remain regular and that the remnant mass stays finite and positive. Hence, the upper bound on the Bertotti-Robinson magnetic parameter is determined not by entropy positivity, but by the regularity of the geometry and the physical behavior of the extremal remnant.
\end{comment}
Using the definitions of $P_0$ and $\delta$, one finds the exact
factorization
\begin{equation}
P_0
=
\left(1-B^2a^2\right)
\left[
1+\frac{B^2m^2}
{\left(1-\frac{1}{2}B^2a^2\right)^2}
\right].
\end{equation}
The factor in square brackets is positive wherever the metric
parameters are well defined. Consequently, positivity of the horizon
entropy requires
\begin{equation}
P_0>0
\quad\Longleftrightarrow\quad
1-B^2a^2>0
\quad\Longleftrightarrow\quad
|Ba|<1.
\end{equation}
The same condition also keeps the extremal radius and remnant mass
finite. Thus, entropy positivity and extremal regularity lead to the
same admissible bound.
\section{Komar mass and integral associated with the horizon generator}
\label{sec:komar}

The Komar construction provides a geometrical way to associate a conserved charge with a Killing symmetry of a spacetime. In a stationary spacetime, the relevant symmetry for the mass is time translation. Therefore, the Komar mass is defined in terms of the timelike Killing vector. In the Kerr-Bertotti-Robinson spacetime, the metric coefficients are independent of both $t$ and $\phi$. Hence, the spacetime is stationary and axisymmetric, and it admits the Killing vectors \cite{62,63,64}
\begin{equation}
	\xi^\mu=\left(\frac{\partial}{\partial t}\right)^\mu=(1,0,0,0),
	\qquad
	\psi^\mu=\left(\frac{\partial}{\partial \phi}\right)^\mu=(0,0,0,1).
\end{equation}
Here $\xi^\mu$ is the stationary Killing vector and $\psi^\mu$ is the axial Killing vector. The Komar mass associated with the stationary Killing vector is defined on a closed two-surface $S$ by
\begin{equation}
	M_K(S)
	=
	-\frac{1}{8\pi}
	\int_S \nabla^\mu \xi^\nu\, dS_{\mu\nu}.
\end{equation}
The surface $S$ is chosen as a two-dimensional surface at fixed $t$ and fixed $r$. Thus, the integration variables on $S$ are $\theta$ and $\phi$. 

The antisymmetric surface element is \cite{64} 
\begin{equation}
	dS_{\mu\nu}
	=
	\left(n_\mu\sigma_\nu-n_\nu\sigma_\mu\right)dA,
\end{equation}
where $n_\mu$ is the future-directed timelike unit normal to the hypersurface $t=\mathrm{constant}$, $\sigma_\mu$ is the outward spacelike unit normal to the surface $r=\mathrm{constant}$, and $dA$ is the induced area element on $S$. Since the two normal directions are the $t$- and $r$-directions, the nonzero components of $dS_{\mu\nu}$ are the mixed $tr$ components. With a definite orientation, one may write
\begin{equation}
	dS_{tr}
	=-dS_{rt}=
	-\sqrt{-g}\,d\theta d\phi,
\end{equation}
For the present geometry, the determinant is $\sqrt{-g}
=
\frac{\Sigma\sin\theta}{P_{0}\Omega^4}.$ The timelike normal one-form is proportional to the gradient of $t$. Therefore,
\begin{equation}
	n_\mu
	=
	-N\partial_\mu t
	=
	(-N,0,0,0),
	\qquad
	N
	=
	\frac{1}{\sqrt{-g^{tt}}},
\end{equation}
where $N$ is fixed by the normalization condition $n_\mu n^\mu=-1$. Although $n_\phi=0$, the contravariant vector $n^\mu$ generally has a $\phi$-component, because the metric is rotating. Indeed,
\begin{equation}
	n^\mu
	=
	g^{\mu\nu}n_\nu
	=
	-Ng^{\mu t}
	=
	(-Ng^{tt},0,0,-Ng^{t\phi}).
\end{equation}
This is the place where the off-diagonal component $g_{t\phi}$ affects the normal vector. The radial unit normal is
\begin{equation}
	\sigma_\mu
	=
	(0,\sqrt{g_{rr}},0,0),
	\qquad
	\sigma_\mu\sigma^\mu=1.
\end{equation}

The metric components needed in the Komar integral are obtained directly from the line element. The corresponding inverse components that enter the Komar mass are
\begin{equation}
	g^{rr}
	=
	\frac{\Omega^2Q}{\Sigma}, \qquad 	g^{tt}
	=
	\frac{\Omega^2}{\Sigma}
	\left[
	\frac{a^2\sin^2\theta}{P}
	-
	\frac{(r^2+a^2)^2}{Q}
	\right],\qquad g^{t\phi}
	=
	\frac{aP_{0}\Omega^2}{\Sigma}
	\left[
	\frac{1}{P}
	-
	\frac{r^2+a^2}{Q}
	\right].
\end{equation}
Using these components, the contraction appearing in the Komar integral becomes
\begin{equation}
	\nabla^\mu \xi^\nu dS_{\mu\nu}
	=
	\sqrt{-g}\,g^{rr}
	\left[
	g^{tt}\partial_r g_{tt}
	+
	g^{t\phi}\partial_r g_{t\phi}
	\right]
	d\theta d\phi .
\end{equation}
This form follows from the fact that $\xi^\mu=(1,0,0,0)$, while the lowered Killing one-form is
\begin{equation}
	\xi_\mu
	=
	g_{\mu\nu}\xi^\nu
	=
	(g_{tt},0,0,g_{t\phi}).
\end{equation}
Therefore, both $\partial_r g_{tt}$ and $\partial_r g_{t\phi}$ contribute to the Komar integral. The second term is absent only in a static spacetime with $g_{t\phi}=0$.

Hence, the finite-radius Komar mass can be written as
\begin{equation}
	M_K(r)
	=
	\frac{1}{8\pi P_{0}}
	\int_0^{2\pi}\int_0^\pi
	\frac{Q\sin\theta}{\Omega^2}
	\left[
	g^{tt}\partial_r g_{tt}
	+
	g^{t\phi}\partial_r g_{t\phi}
	\right]
	d\theta d\phi .
\end{equation}
The above expression defines a finite-surface Komar charge. It should be interpreted with care. Since the Kerr-Bertotti-Robinson spacetime is supported by a magnetic background, the geometry is not asymptotically flat in the ordinary Kerr sense. Therefore, the Komar integral at finite $r$ measures not only the black hole contribution, but also the gravitational contribution of the magnetic background inside the integration surface.

In the weak-field regime, $B^2m^2\ll 1,$ the Komar mass can be expanded as
\begin{equation}
	M_K(r)
	=
	M_K^{(0)}(r)+B^2M_K^{(2)}(r)+O(B^4).
\end{equation}
The zeroth-order term is the usual Kerr result, $M_K^{(0)}(r)=m.$ Keeping the leading magnetic correction, one finds
\begin{equation}
	M_K(r)
	=
	m+\frac{B^2}{4}\mathcal I_2(r)+O(B^4),
\end{equation}
where
\begin{equation}
	\begin{aligned}
		\mathcal I_2(r)
		={}&
		\frac{1}{6a^3(r^2+a^2)}
		\Bigg\{
		8a^3(r^2+a^2)
		\left[
		-4a^2m+a^2r+5m^2r-3mr^2+r^3
		\right]
		\\
		&\quad
		+24amr^3(r^2+a^2)(m-r)
		\\
		&\quad
		+12am
		\left[
		-a^6-3a^4r^2
		+3a^2mr^3
		-4a^2r^4
		+3mr^5
		-2r^6
		\right]
		\\
		&\quad
		-12mr^2(r^2+a^2)^2(5m-4r)
		\tan^{-1}\left(\frac{a}{r}\right)
		\Bigg\}.
	\end{aligned}
\end{equation}
This result explicitly shows that, for $B\neq0$, the Komar charge depends on the radius of the integration surface. This behavior is expected because the magnetic background carries energy and contributes to the gravitational charge.

Several limiting cases are useful, e.g., in the Kerr limit, $M_K(r)=m$. Therefore, the mass parameter coincides with the Komar mass in the usual asymptotically flat Kerr spacetime. In the nonrotating limit $a\to0$, the weak-field expression has a regular limit,
\begin{equation}
	\mathcal I_2(r)\big|_{a=0}
	=
	\frac{4}{3}r^3-\frac{8}{3}mr^2,
\end{equation}
and hence
\begin{equation}
	M_K(r)\big|_{a=0}
	=
	m
	+
	B^2
	\left(
	\frac{r^3}{3}
	-
	\frac{2mr^2}{3}
	\right)
	+O(B^4).
\end{equation}
If also $m=0$, one obtains the leading pure magnetic-background contribution $	M_K(r)\big|_{m=a
	=0}
=
\frac{B^2r^3}{3}
+O(B^4).$ It is also useful to introduce the Komar integral associated with the horizon generator. The event horizon is generated by
\begin{equation}
	\chi^\mu
	=
	\xi^\mu+\Omega_h\psi^\mu
	=
	(1,0,0,\Omega_h),
\end{equation}
where $\Omega_h$ is the angular velocity of the horizon. The corresponding Komar charge is
\begin{equation}
	K_\chi
	=
	-\frac{1}{8\pi}
	\int_H
	\nabla^\mu \chi^\nu dS_{\mu\nu}.
\end{equation}
For a Killing horizon, this charge is related to the surface gravity and the horizon area by Eqs. \eqref{eq:temprature}-\eqref{eq:entropy}
\begin{equation}
	K_\chi
	=
	\frac{\kappa_h A_h}{4\pi}
	=
	2T_hS_h=
	\frac{
		\gamma r_h^2-a^2\zeta_h
	}
	{
		P_{0}r_h\left(
		\gamma+\zeta_h
		\right)
	}.
\end{equation}

This quantity is the Komar charge of the horizon generator, not the total mass associated with the Killing vector $\xi^\mu=\partial_t^\mu$. In the Kerr limit, $K_\chi=
\frac{r_h^2-a^2}{2r_h}.$ Using $r_h=m+\sqrt{m^2-a^2}$, this becomes $K_\chi^{\rm Kerr}
=
\sqrt{m^2-a^2}.$ Thus, for rotating Kerr, $	K_\chi^{\rm Kerr}\neq m.$

           \section{Photon Motion}
\label{sec:photon-motion}

            We now study the motion of photons in the Kerr-Bertotti-Robinson geometry. Since the black hole shadow and the photon region are determined by unstable null geodesics, the Hamilton-Jacobi formalism provides a convenient method for deriving the corresponding equations of motion.
			
			\begin{equation}
				\frac{1}{2}g^{\mu\nu}p_\mu p_\nu
				=0 .
			\end{equation}
			
			The Hamilton-Jacobi action can be written as \cite{Hassanabadi2019, Li2025a, Li2025b}
			
			\begin{equation}
				S=
				\frac{1}{2}\mu^2\lambda
				-Et+L_z\phi+S_r(r)+S_\theta(\theta),
			\end{equation}
			
			where $\lambda$ is the affine parameter, $E=-p_t$ is the conserved energy, $L_z=p_\phi$ is the conserved angular momentum along the symmetry axis, and $\mu=0$ for photons.
			
Using the metric functions $P$, $Q$, $\Sigma$, $\Omega$, one obtains the radial potential
			\begin{equation}
				R(r) = \left[(r^2 + a^2)E - a P_{0}L_z\right]^2 - Q\mathcal{K},
				\label{eq:radial_potential}	\end{equation}
			and the angular potential
			\begin{equation}
		\Theta(\theta)=	P(\theta)\mathcal{K}-	\left(
				aE\sin\theta-\frac{P_{0}L_z}{\sin\theta}
				\right)^2 ,
			\end{equation}
			
			where $\mathcal{K}$ is the separation constant. The conformal factor $\Omega^{-2}$ does not affect the unparametrized null geodesic paths; it only changes the affine parametrization along the photon trajectory.
			
			\begin{equation}
				\begin{split}
				R(r) &= \left[(r^2 + a^2) - aP_{0}\xi\right]^2 - Q\eta,\\
				\Theta(\theta)&=	P(\theta)\eta-
				\left(
				a\sin\theta-\frac{\xi P_{0}}{\sin\theta}
				\right)^2.
				\end{split}
			\end{equation}
			
	%\begin{equation}
	%			\frac{a}{Q(r)}
	%			\left[(r^2+a^2)E-aL_z\right]
%				+
	%			\frac{1}{P(\theta)\sin^2\theta}
%				\left(L_z-aE\sin^2\theta\right).
%			\end{equation}
			
			The functions $R(r)$ and $\Theta(\theta)$ determine the allowed regions of photon motion. In particular, the radial motion is allowed only where $				R(r)\geq0,$ while the angular motion is allowed only where $	\Theta(\theta)\geq0.$

			The zeros of $R(r)$ correspond to radial turning points, whereas the zeros of $\Theta(\theta)$ determine the angular turning points of the photon trajectory. Unstable spherical photon orbits arise when the radial potential satisfies both $R(r)=0$ and $R'(r)=0$. These orbits form the photon region and provide the fundamental input for constructing the black hole shadow in the following section.

            \section{Spherical Photon Orbits and Celestial Coordinates}
\label{sec:photon-orbits}
The boundary of the black hole shadow is determined by unstable spherical photon orbits \cite{44}. These are null geodesics for which the radial coordinate remains constant, while the photon can still move in the angular direction. Therefore, the spherical photon orbits are obtained by imposing the conditions \cite{66}

\begin{equation}
	R(r_p)=0,
	\qquad
	\frac{dR(r)}{dr}\bigg|_{r=r_p}=0,
\end{equation}

where $r_p$ denotes the radius of the spherical photon orbit, whereas,
\begin{subequations}
    \begin{align}
       &\xi(r) = \frac{1}{P_{0}}[\frac{r^2 + a^2}{a} - \frac{4rQ}{aQ'}],\label{eq:xir}\\
       &\eta(r) = \frac{16r^2Q}{(Q')^2}.\label{eq:etar}
    \end{align}
\end{subequations}
The stability of these spherical photon orbits is determined by the second derivative of the radial potential $R(r)$. With the convention $R\ge0$, instability is identified by $R''(r_p)>0$. This condition ensures that any small perturbation will cause the photon to either fall into the black hole or escape to infinity. Only these unstable orbits contribute to the boundary of the black hole shadow, as stable orbits would not be visible to a distant observer. The shadow boundary is therefore constructed by selecting the subset of spherical photon orbits that satisfy this instability condition.

These two functions characterize the photon region and determine the apparent boundary of the shadow, having

\begin{subequations}
	\begin{align}
    &\hat{R}_o = \left[(r_o^2 + a^2) - aP_{0}\xi\right]^2 - Q_o\eta,\\ 
	&\hat{\Theta}_o = P_o\eta - \left(a\sin\theta_o - \frac{\xi P_{0}}{\sin\theta_o}\right)^2,
    \end{align}
\end{subequations}

We also define $	F_o=Q_o-a^2P_o\sin^2\theta_o .$

The condition $F_o>0$ ensures that the static observer is located outside the stationary-limit surface, so that,

\begin{subequations}
	\begin{align}
	&\alpha = -r_o \frac{\sqrt{F_o}}{\sin \theta_o \sqrt{P_o \hat{R}_o}} \left[ P_{0}\xi - \frac{a \sin^2 \theta_o (Q_o - P_o (r_o^2 + a^2))}{F_o} \right],\label{eq:alpha}\\
	&\beta = \pm r_o \sqrt{\frac{\hat{\Theta}_o Q_o}{P_o \hat{R}_o}}.\label{eq:beta}
    \end{align}
\end{subequations}

Here, the plus and minus signs correspond to the upper and lower parts of the shadow boundary, respectively.

The shadow curve is obtained by substituting $\xi(r_p)$ and $\eta(r_p)$ into the celestial coordinates and varying $r_p$ over the allowed range of unstable spherical photon orbits. In this way, each spherical photon orbit is mapped to a point on the observer's celestial plane. The collection of these points forms the apparent boundary of the Kerr-Bertotti-Robinson black hole shadow.
 For a static observer at a finite radius $r_o$, the celestial coordinates are given by Eqs.\eqref{eq:alpha} and \eqref{eq:beta} \cite{67,68,69}. 
For a finite-radius observer satisfying
\begin{equation}
\frac{m}{r_o}\ll1,\qquad Br_o\ll1,
\end{equation}
the exact expressions admit a perturbative expansion in the small parameters $m/r_o$ and $Br_o$, yielding the following approximate forms. This should not be interpreted as the standard asymptotic observer limit, since the Kerr-Bertotti-Robinson spacetime is neither asymptotically flat nor asymptotically AdS. Instead, the approximation is valid in the intermediate radial regime $m\ll r_o\ll B^{-1}$, where the observer is sufficiently far from the black hole while remaining well within the characteristic magnetic length scale. Then
\begin{equation}
	\alpha \simeq -\frac{P_{0}\xi}{\sin\theta_o}, \qquad
	\beta \simeq \pm \sqrt{\eta - \left(a\sin\theta_o - \frac{P_{0}\xi}{\sin\theta_o}\right)^2}
\end{equation}

The deviations from these expressions are controlled by the Bertotti-Robinson corrections in the metric functions $P(\theta)$ and $Q(r)$, and are of order $B^2r_o^2$, $B^2m^2$, and $B^2a^2$.

For each polar angle $\theta$, solving  $\Theta(\theta)=0=0$ with $\xi(r)$ and $\eta(r)$ from Eqs. \eqref{eq:xir}-\eqref{eq:etar} gives the limiting radii $r_{ph}^-(\theta)$ and $r_{ph}^+(\theta)$ \cite{16,67}. The photon region at that angle then lies between them:

\begin{equation}
r_{\rm ph}^{-}(\theta) \le r \le r_{\rm ph}^{+}(\theta).
\end{equation}

At the equatorial plane $\theta = \pi/2$, we have $P=1$ and $\sin\theta=1$. Then the photon-region condition reduces to
%\begin{equation}
%	\eta(r) - (a - \xi(r))^2 = 0.
%\end{equation}
\begin{equation}
    \eta(r)-\left[a-P_{0}\xi(r)\right]^2=0.
\end{equation}
The factor $P_{0}$ is required by the regular-axis azimuthal
normalization and is consistent with the angular potential
$\Theta(\theta)$. The two solutions of this equation are the prograde and retrograde equatorial photon radii.
The photon sphere radii $r_{ph}^\pm$ (where the superscripts denote prograde and retrograde branches, respectively) are determined by the condition $\Theta_\theta(r_p) = 0$, where we have defined $\Theta_{\theta}(r) \equiv \eta(r) - [a - P_{0}\xi(r)]^2$ and $b=B m$. In dimensionless form, we find
\begin{equation}
	\frac{r_{ph}^{\pm}}{m}=x_\pm
	+
	b^2
	\frac{x_\pm^2(11-2x_\pm+3x_\pm^2)}{24}
	+O(b^4).
	\label{eq:rp_general}
\end{equation}
with
\begin{equation}
	x_{\pm} = 2\left[1 + \cos\left(\frac{2}{3}\cos^{-1}(\pm j)\right)\right].
\end{equation}
For $a>0$, i.e., $j>0$, $x_-$ corresponds to the prograde branch and $x_+$ to the retrograde branch.
Equivalently, in dimensional form,

\begin{equation}\label{eq:photonradii}
	\begin{aligned}
		r_{\rm ph}^{\pm}(\theta)
		&\simeq 3m \pm \frac{2a}{\sqrt{3}}\sin\theta - \frac{2a^2}{9m}(2 - \sin^2\theta) \\
		&\quad
        + B^2[12m^3 \pm (\sqrt3/3)m^2a\sin\theta(29-\sin^2\theta) + (ma^2/18)(-8\sin^4\theta+156\sin^2\theta-49)].
	\end{aligned}
\end{equation}
	\begin{figure}[htbp]
	\centering
	\subfigure[]{
		\begin{minipage}[t]{0.5\textwidth}
			\centering
			\includegraphics[width=\linewidth]{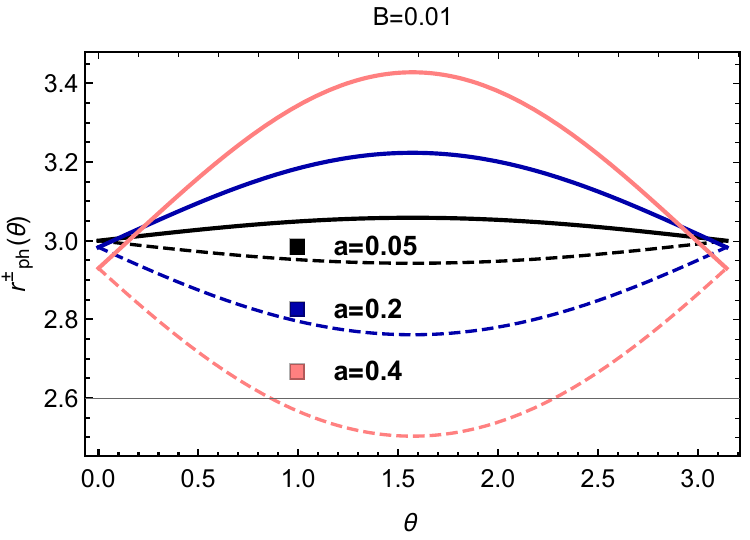}
			%\caption{Real part}
	\end{minipage}}%
	\subfigure[]{
		\begin{minipage}[t]{0.5\linewidth}
			\centering
			\includegraphics[width=\linewidth]{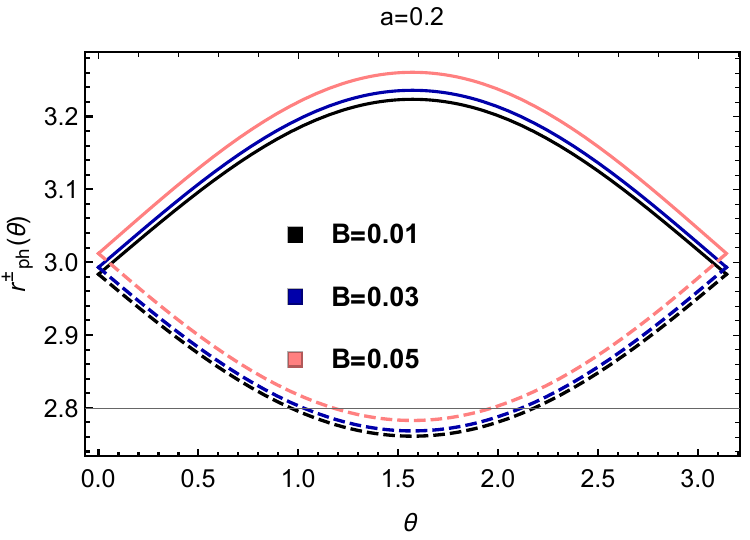}
			%\caption{Imaginary part
			%}
	\end{minipage}}%
	\centering
	\caption{ \small Angular dependence of the photon-region boundaries $r_{\rm ph}^{\pm}(\theta)$. Panel (a) shows $r_{\rm ph}^{\pm}(\theta)$ for fixed $B=0.01$ and different values of the rotation parameter. The solid curves represent the retrograde branch $r_{\rm ph}^{+}$, while the dashed curves represent the prograde branch $r_{\rm ph}^{-}$. Panel (b) shows the same photon-region boundaries for fixed $a=0.2$ and different values of $B$. The curves are obtained from the weak-field and slow-rotation expansion in Eq. \eqref{eq:photonradii}.
	}\label{fig:rph}
\end{figure}

Figure \eqref{fig:rph} illustrates the angular structure of the photon region. In the nonrotating limit, the prograde and retrograde photon branches coincide, but rotation breaks this degeneracy through frame dragging. As the spin increases, photons moving in the prograde direction can orbit closer to the black hole, while retrograde photons are pushed to larger radii. This explains the widening gap between the dashed and solid curves in panel (a). The angular dependence is controlled mainly by the $\sin\theta$ terms in the photon-radius expansion, so the splitting is strongest near the equatorial plane and becomes weaker toward the poles. Panel (b) shows that increasing the Bertotti-Robinson parameter shifts the photon-region boundaries outward. Therefore, the external background changes the overall radial location of the photon region, while rotation mainly controls the prograde-retrograde asymmetry.

In the equatorial plane, $\theta = \pi/2$, this becomes, with $r_{ph}^-$ denoting the prograde branch and $r_{ph}^+$ the retrograde branch for $a > 0$:
\begin{equation}
	r_{\text{ph}}^{\pm}\left(\frac{\pi}{2}\right) \simeq 3m \pm \frac{2a}{\sqrt{3}} - \frac{2a^2}{9m} + B^2 \left[12m^3 \pm \frac{28\sqrt{3}}{3} m^2 a + \frac{11}{2} ma^2\right].
\end{equation}
These expressions reveal that the Bertotti-Robinson background shifts both photon endpoints outward at leading order. The spin $a$ contributes through odd powers in the individual endpoints, whereas the total shadow area depends solely on even powers of $a$, as we shall verify below.

			\section{Ergosphere Structure}
\label{sec:ergosphere}

The ergosphere provides a direct geometric measure of the rotational properties of black holes, thus playing an important role in energy extraction processes. Hence, in the Kerr-Bertotti-Robinson geometry, this region is modified through both the radial and angular metric functions. Accordingly, in this section, we analyze how the parameter $B$ deforms the stationary limit surface and changes the size of the ergoregion. We first derive the equation that determines the ergosurface and discuss its Kerr limit. We then perform a weak-field expansion in order to isolate the leading Bertotti-Robinson corrections to the angular profile of the ergosphere. Finally, we introduce the ergosphere thickness and the photon-ergosphere gap, quantifying respectively the radial extension of the energy-extraction region and its separation from the photon-region boundaries.

            \subsection{Ergosurface Equation}
			
            The ergosphere is a characteristic feature of rotating black holes. It is the region outside the event horizon where the stationary Killing vector $\partial_t$ becomes spacelike, so that no observer can remain static with respect to infinity. The boundary of this region is called the stationary-limit surface or ergosurface, and it is determined by the condition 				$g_{tt}=0 $ \cite{44}.
Using Eq.\eqref{eq-metric}
			\begin{equation}
				g_{tt}=\frac{1}{\Omega^2\Sigma}
				\left[
				a^2P(\theta)\sin^2\theta-Q(r)
				\right].
			\end{equation}
			
			Since the conformal factor $\Omega^2$ and $\Sigma$ do not vanish on a regular stationary-limit surface, the ergosurface condition reduces to $Q(r)=a^2P(\theta)\sin^2\theta .$ Substituting the explicit forms of $P$ and $Q$, one obtains

			\begin{equation}
				(1 + B^2r^2)\left[\left(1 - B^2m^2\delta\right)r^2 - 2m\gamma r + a^2\right] - a^2\sin^2\theta\left[1 + B^2\left(m^2\delta - a^2\right)\cos^2\theta\right] = 0.
			\end{equation}
			This equation determines the angular-dependent ergosurface radius, $r=r_{\rm B,ergo}^+(\theta).$
			
			The outer solution, denoted by $r_{\rm B,ergo}^+(\theta)$, defines the boundary of the physical ergoregion. At the poles, where $\sin\theta=0$, the ergosurface coincides with the event horizon. Away from the poles, the rotational term proportional to $a^2\sin^2\theta$ shifts the stationary-limit surface outside the horizon, producing a finite ergoregion. The Bertotti-Robinson parameter modifies this structure through both $Q(r)$ and $P(\theta)$, and therefore changes the radial and angular shape of the ergosphere.
			\subsection{Kerr Limit and Weak-Field Expansion}
%The ergosphere plays an important role in the physics of rotating black holes because it is the region where rotational energy extraction becomes possible. In this region, the timelike Killing vector becomes spacelike, and particles may have negative energy with respect to a distant observer. As a result, the ergosphere is closely related to processes such as the Penrose mechanism, superradiant scattering, and the extraction of rotational energy from the black hole. Therefore, understanding how the Bertotti-Robinson parameter modifies the ergosphere is important for both geometrical and physical reasons.
Having obtained the ergosurface equation and its Kerr limit, we now examine how the Bertotti-Robinson background modifies its angular profile.
\begin{equation}
	r_{0,\mathrm{ergo}}^+(\theta)=m+\sqrt{m^2-a^2\cos^2\theta},
	\label{eq:kerr_ergo}
\end{equation}

which defines the physical boundary of the Kerr ergosphere. It coincides with the event horizon at the poles, $\theta=0,\pi$, and reaches its maximum radial extension at the equatorial plane, $\theta=\pi/2$.

\begin{equation}
	r_{B,\mathrm{ergo}}^+(\theta)=r_{0,\mathrm{ergo}}^+(\theta)
	+
	B^2 r_{1,\mathrm{ergo}}^+(\theta)
	+
	\mathcal{O}(B^4).
	\label{eq:ergo_expansion}
\end{equation}

Here $r_{0,\mathrm{ergo}}^+(\theta)$ is the Kerr outer ergosurface, while $r_{1,\mathrm{ergo}}^+(\theta)$ gives the leading correction induced by the Bertotti-Robinson background.

\begin{equation}
	\begin{split}
		r_{1,\mathrm{ergo}}^+(\theta)
		=
		-\frac{
			a^2\left(r_{0,\mathrm{ergo}}^+\right)^2\sin^2\theta
			-m^2\left(r_{0,\mathrm{ergo}}^+\right)^2
			+ma^2 r_{0,\mathrm{ergo}}^+
			-a^2\sin^2\theta\,(m^2-a^2)\cos^2\theta
		}{
			2\left(r_{0,\mathrm{ergo}}^+-m\right)
		}.
	\end{split}
	\label{eq:ergo_correction}
\end{equation}
%Unless otherwise stated, the plotted analytic quantities are evaluated in the week-field and small-spin approximation.
This expression shows that the Bertotti-Robinson background modifies the Kerr ergosphere only at order $B^2$. The correction $r_{1,\mathrm{ergo}}^+(\theta)$ depends explicitly on the polar angle, which means that the magnetic deformation changes not only the overall size of the ergosphere but also its angular profile. In particular, the deformation affects the equatorial and near-polar regions differently, leading to a modified distribution of the rotational energy-extraction zone around the black hole.

\begin{figure}[htbp]
	\centering
	\subfigure[]{
		\begin{minipage}[t]{0.5\textwidth}
			\centering
			\includegraphics[width=\linewidth]{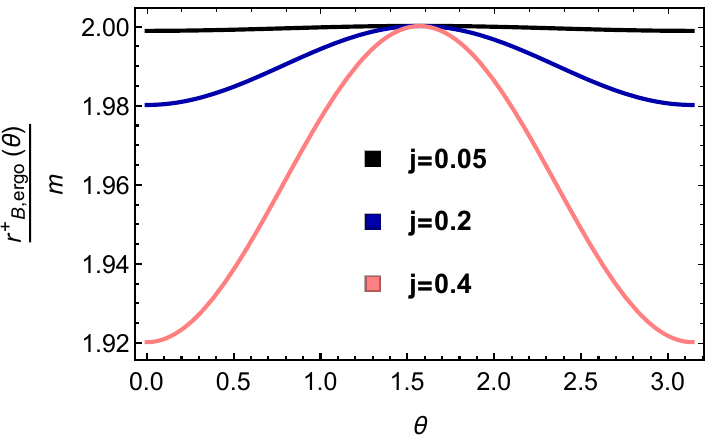}
			%\caption{Real part}
	\end{minipage}}%
	\subfigure[]{
		\begin{minipage}[t]{0.5\linewidth}
			\centering
			\includegraphics[width=\linewidth]{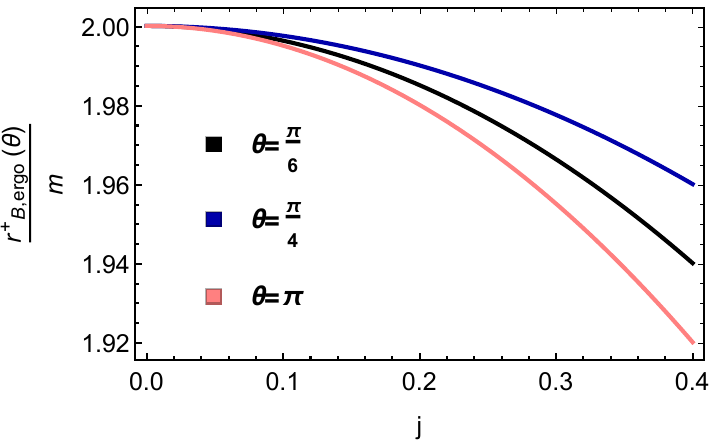}
			%\caption{Imaginary part
			%}
	\end{minipage}}%
	\centering
	\caption{ \small Ergosurface radius obtained from the combined weak-field and
		slow-rotation approximation in Eq.\eqref{eq:rergospher}. Panel (a) shows the angular
		profile $r^+_{B,\rm ergo}(\theta)/m$ for different spin parameters
		$j=a/m$. Panel (b) shows $r^+_{B,\rm ergo}/m$ as a function of $j$
		for fixed polar angles. The plots are intended only within the
		perturbative regime $b=Bm\ll1$ and $j\ll1$.
	}\label{fig:rergo}
\end{figure}

Having obtained the explicit form of the ergosphere radius, we now examine its angular dependence. Using the weak-field expansion for the ergosphere radius, we have to leading order in the spin and magnetic field:
\begin{equation}\label{eq:rergospher}
	r_{\rm B,ergo}^+(\theta) \simeq 2m - \frac{a^2}{2m}\cos^2\theta
	+ B^2\left[2m^3 - \frac{m a^2}{2}(\cos^4\theta - 5\cos^2\theta + 6)\right].
\end{equation}
Figure \eqref{fig:rergo} (a) shows the angular dependence of the ergosurface radius
	$r^+_{B,\rm ergo}(\theta)/m$ for fixed values of the spin parameter
	$j=a/m$. The profile is symmetric about the equatorial plane and
	reaches its maximum near $\theta=\pi/2$. As the spin increases, the
	angular deformation becomes stronger, and the polar parts of the
	ergosurface move inward.
	
Figure \eqref{fig:rergo} (b) shows $r^+_{B,\rm ergo}(\theta)/m$ as a function of the spin
	parameter $j$ for fixed polar angles. The radius decreases with
	increasing $j$, and the decrease is strongest near the polar direction.

 Differentiation gives
\begin{equation}
	\frac{dr_{\rm B,ergo}^+(\theta)}{d\theta}
	= a^2 \sin\theta \cos\theta \left[\frac{1}{m} - B^2 m (5 - 2\cos^2\theta)\right].
\end{equation}

The extrema satisfy $dr_{\rm B,ergo}^+(\theta)/d\theta = 0$. This gives three possibilities:
\begin{itemize}
	\item[-] $\sin\theta= 0$, which corresponds to the poles $\theta = 0, \pi$;
	\item[-] $\cos\theta = 0$, which corresponds to the equator $\theta = \pi/2$;
	\item[-] the bracket vanishes: $\frac{1}{m} - B^2 m (5 - 2\cos^2\theta) = 0$, which gives
	\begin{equation}
		\cos^2\theta = \frac{1}{2}\left(5 - \frac{1}{B^2 m^2}\right).
	\end{equation}
\end{itemize}

In the weak-field regime $B^2 m^2 \ll 1$, the quantity $5 - 1/(B^2 m^2)$ is negative, so the third solution is unphysical. Thus, within the perturbative regime, the only relevant extrema are the poles and the equator.

At the poles, $\theta = 0, \pi$, so  substituting into the ergosphere radius gives
\begin{equation}\label{eq:rhorizon2}
	r_{\rm B,ergo}^{+}(0)=r_{\rm B,ergo}^{+}(\pi) \simeq 2m - \frac{a^2}{2m} + B^2(2m^3 - m a^2).
\end{equation}
\subsection{Ergosphere Thickness}

The local thickness of the ergosphere is defined as the radial separation between the
outer stationary-limit surface and the outer event horizon,
\begin{equation}
	T_{\rm B,ergo}(\theta)=r_{\rm B,ergo}^{+}(\theta)-r_{+}.
\end{equation}
Here $r_{\rm B,ergo}^{+}(\theta)$ denotes the largest positive root of the stationary-limit
equation, 
whereas the horizon radius is obtained from $Q(r_{+})=0$, or equivalently
$\Delta(r_{+})=0$. 
The outer horizon is given by Eq. \eqref{eq:rhorizon}. The corresponding dimensionless ergosphere thickness is
\begin{equation}
	\bar T_{\rm B,ergo}(\theta)
	=
	\frac{r_{\rm ergo}^{+}(\theta)-r_{+}}{r_{+}} .
\end{equation}

In the Kerr limit $B=0$, Eq. \eqref{eq:kerr_ergo} is yielded. The thickness vanishes at the poles and reaches its maximum at the equatorial plane.

Hence
\begin{equation}
	T_{\rm B,ergo}(\theta)
	=
	T_{\rm ergo}^{\rm Kerr}(\theta)
	+
	B^{2}
	\left[
	r^+_{1,\rm ergo}(\theta)-r_{+}^{(1)}
	\right]
	+
	\mathcal{O}(B^{4}).
\end{equation}

To characterize the global size of the ergosphere, we define the angularly averaged
thickness
\begin{equation}
	\langle T_{\rm B,ergo}\rangle
	=
	\frac{1}{\pi}
	\int_{0}^{\pi}
	\left[
	r_{\rm B,ergo}^{+}(\theta)-r_{+}
	\right]d\theta .
\end{equation}
Introducing the dimensionless spin $j=\frac{a}{m}$,  $\kappa=\sqrt{1-j^{2}}$, and the complete elliptic integrals
\begin{equation}
	K(j)=\int_{0}^{\pi/2}\frac{d\chi}{\sqrt{1-j^{2}\sin^{2}\chi}},
	\qquad
	E(j)=\int_{0}^{\pi/2}\sqrt{1-j^{2}\sin^{2}\chi}\,d\chi ,
\end{equation}
the Kerr contribution becomes
\begin{equation}
	\frac{\langle T_{\rm ergo}\rangle_{\rm Kerr}}{m}
	=
	\frac{2}{\pi}E(j)-\kappa .
\end{equation}
The full weak-field result can then be written as
\begin{equation}\label{eq:TBergo}
	\frac{\langle T_{\rm B,ergo}\rangle}{m}
	=
	\frac{2}{\pi}E(j)-\kappa
	+
	(Bm)^{2}\,\mathcal{D}_{\rm ergo}(j)
	+
	\mathcal{O}\!\left((Bm)^{4}\right),
\end{equation}
where the Bertotti-Robinson correction is
\begin{equation}
\mathcal{D}_{\mathrm{ergo}}(j) = 
\frac{1}{	3\pi j^2}\Big[
	2(1 - 2j^2)\,E(j) - (9j^4 - 11j^2 + 2)\,K(j) + 3\pi j^2 (1 - j^2)\Big]- \frac{1}{2} (1 + \kappa)^2.
\end{equation}
This expression gives the analytic solution of the integral appearing in the averaged
thickness at order $B^{2}$. The limit $j\to0$ is understood by continuous expansion.

\begin{figure}[htbp]
	\centering
	
	\includegraphics[width=.750\textwidth]{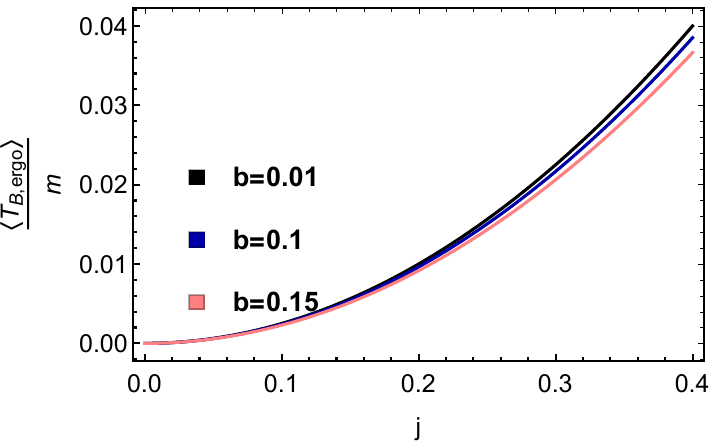}
	
	\caption{Dimensionless averaged ergosphere thickness $ \langle T_{\text{ergo}} \rangle/m $ as a function of the spin parameter $ j = a/m $ for different values of $ b = Bm $. The thickness vanishes in the nonrotating limit and grows rapidly with spin. Increasing $ b $ slightly decreases the thickness, showing that the Bertotti-Robinson background weakly suppresses the rotational energy-extraction region. The curves are obtained from the $\mathcal{O}(B^2)$ weak-field expansion in Eq. \eqref{eq:TBergo}.}
	\label{fig:Tergo}
\end{figure}

Figure \eqref{fig:Tergo} presents the dimensionless ergosphere thickness $T_{\rm ergo}/m$ as a function of the spin parameter $j$. The thickness vanishes in the nonrotating limit, as expected, because the ergosurface coincides with the event horizon when $a=0$. As the spin increases, the ergosphere expands and the thickness grows rapidly. For fixed $j$, increasing $b=Bm$ slightly decreases the thickness, in agreement with the analytic weak-field result that the Bertotti-Robinson background gives a negative correction to the averaged ergosphere thickness. Therefore, rotation enlarges the energy-extraction region, while the external Bertotti-Robinson background weakly suppresses it.

Thus, to leading order in the spin,
\begin{equation}
	\langle T_{\rm ergo}\rangle
	\simeq
	\frac{a^{2}}{4m}
	\left[
	1-\frac{15}{4}B^{2}m^{2}
	\right].
\end{equation}
The Bertotti-Robinson background therefore, decreases the averaged ergosphere thickness
at leading order in $B^{2}$. At the equatorial plane, the maximum thickness is $T_{\rm ergo}^{\rm max}
=
T_{\rm ergo}\left(\frac{\pi}{2}\right).$

Using the same weak-field expansion, one obtains
\begin{equation}
	\frac{T_{\rm ergo}^{\rm max}}{m}
	=
	1-\kappa
	+
	b^{2}
	\left(
	1-\kappa-\frac{5}{2}j^{2}
	\right)
	+
	\mathcal{O}\!\left(b^{4}\right).
\end{equation}
For small spin,
\begin{equation}
	\frac{T_{\rm ergo}^{\rm max}}{m}
	=
	\frac{j^{2}}{2}
	+
	\frac{j^{4}}{8}
	-
	b^{2}
	\left[
	2j^{2}
	-
	\frac{j^{4}}{8}
	+
	\mathcal{O}(j^{6})
	\right]
	+
	\mathcal{O}\!\left(b^{4}\right).
\end{equation}
This form correctly gives zero ergosphere thickness in the nonrotating limit $a=0$.			
			\subsection{Photon-ergosphere gap}

We now define the photon-ergosphere gap using the improved photon-region expansion. The photon-region boundaries are obtained from Eq. \eqref{eq:photonradii}.

Here $r_{\rm ph}^{-}$ is the prograde branch and $r_{\rm ph}^{+}$ is the retrograde branch (for $a>0$).

The photon-ergosphere gap is defined as
\begin{equation}
	\Delta_{\rm ph-ergo}^{\pm}
	= \frac{1}{\pi r_+}\int_0^\pi \left[r_{\rm ph}^{\pm}(\theta) - r_{\rm ergo}(\theta)\right]d\theta.
\end{equation}
The lower sign corresponds to the prograde branch, while the upper sign corresponds to the retrograde branch. The ergosphere radius is given by Eq. \eqref{eq:rergospher} to the same order, and the horizon radius by Eq. \eqref{eq:rhorizon}.

\begin{figure}[htbp]
	\centering
	\subfigure[]{
		\begin{minipage}[t]{0.5\textwidth}
			\centering
			\includegraphics[width=\linewidth]{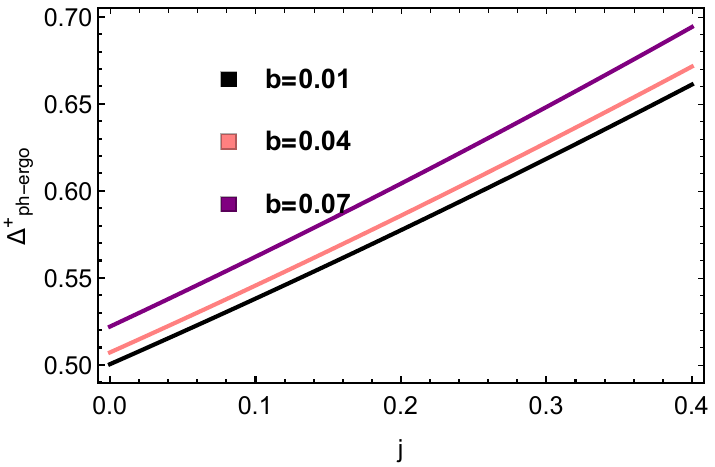}
			%\caption{Real part}
	\end{minipage}}%
	\subfigure[]{
		\begin{minipage}[t]{0.5\linewidth}
			\centering
			\includegraphics[width=\linewidth]{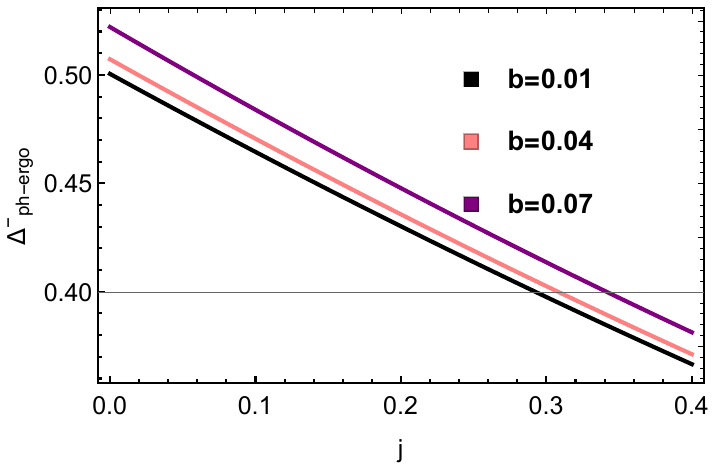}
			%\caption{Imaginary part
			%}
	\end{minipage}}%
	\centering
	\caption{ \small Photon-ergosphere gap as a function of the spin parameter $ j = a/m $ for different values of $ b = Bm $: (a) retrograde branch $ \Delta_{\text{ph-ergo}}^{+} $ and (b) prograde branch $ \Delta_{\text{ph-ergo}}^{-} $. The two branches coincide at $ j = 0 $. As $ j $ increases, frame dragging pushes the retrograde branch outward and pulls the prograde branch inward. Larger $b$ increases the overall separation and enhances the branch asymmetry. The curves are obtained from the combined weak-field and slow-rotation expansion in Eq.~\eqref{eq:deltaphergo}.	}\label{fig:photonergosphergap}
\end{figure}

Therefore, the  photon-ergosphere gap takes the form
\begin{equation}
	\Delta_{\mathrm{ph-ergo}}^{\pm} = 
	\frac{
		\pi m \pm \frac{4a}{\sqrt{3}} - \frac{\pi a^2}{12m} 
		+ B^2 [10\pi m^3 \pm (170\sqrt3/9)m^2a + (487\pi/144)ma^2]
	}{
		\pi \left[ 2m - \frac{a^2}{2m} + B^2 (2m^3 - m a^2) \right]
	}.
\end{equation}
Let us introduce the dimensionless parameters $j = \frac{a}{m}$ and $b=B m$ defined in Sec. \eqref{GeometryandHorizonstructure}. Expanding the previous expression in powers of $j$ and $b$ gives
\begin{equation}\label{eq:deltaphergo}
	\Delta_{\rm ph-ergo}^{\pm}\simeq 1/2 \pm (2\sqrt3/3\pi)j + (1/12)j^2
+ b^2[9/2 \pm (79\sqrt3/9\pi)j + (859/288)j^2].
\end{equation}
Figures \eqref{fig:photonergosphergap} (a) and \eqref{fig:photonergosphergap} (b) show the photon-ergosphere gap as a function of the spin parameter $j$. In the nonrotating limit, the prograde and retrograde branches coincide. As the spin increases, the two branches separate: the retrograde branch increases with $j$, while the prograde branch decreases. This behavior reflects the frame-dragging effect of rotation, which pushes the retrograde photon orbit outward and pulls the prograde photon orbit inward. For larger values of $b=Bm$, the gap shifts upward, indicating that the Bertotti-Robinson background increases the average separation between the photon region and the ergosphere. The mixed spin-background terms enhance the asymmetry between the two branches.

The average of the two branches removes the odd-in-spin terms:
%\begin{equation}
%	\overline{\Delta}_{\rm ph-ergo}
%	= \frac{\Delta_{\rm ph-ergo}^{+} + \Delta_{\rm ph-ergo}^{-}}{2}
%	\simeq \frac{1}{2} + \frac{1}{12}\frac{a^2}{m^2}
%	+ b^2\left[\frac{9}{2} + \frac{851}{288}\frac{a^2}{m^2}\right].
%\end{equation}
\begin{equation}
\overline{\Delta}_{\rm ph-ergo}
=
\frac{\Delta_{\rm ph-ergo}^{+}+\Delta_{\rm ph-ergo}^{-}}{2}
\simeq
\frac{1}{2}
+\frac{1}{12}\frac{a^2}{m^2}
+b^2\left[
\frac{9}{2}
+\frac{859}{288}\frac{a^2}{m^2}
\right].
\end{equation}
The difference between the two branches, which measures the prograde-retrograde splitting, is
\begin{equation}
	\Delta_{\rm ph-ergo}^{+} - \Delta_{\rm ph-ergo}^{-}
	\simeq \frac{4\sqrt{3}}{3\pi}\frac{a}{m}
	+ b^2 \frac{158\sqrt{3}}{9\pi}\frac{a}{m}.
\end{equation}

When rotation is included, the prograde photon region moves closer to the black hole (reducing the gap), while the retrograde branch moves outward (increasing the gap). The $B^2 a$ terms amplify this asymmetry.

\section{Shadow Observables}
\label{sec:shadow-observables}

In this section, we focus on two integrated observables associated with the apparent shadow of the Kerr-Bertotti-Robinson black hole. We first compute the shadow area for an equatorial observer by using the parametric curve determined by the celestial coordinates of unstable spherical photon orbits. We then expand this quantity in the weak-field and small-spin regime in order to isolate the leading correction produced by the external Bertotti-Robinson parameter. Finally, we introduce the magnetic shadow susceptibility, which measures the response of the shadow area to variations of the external background. These observables show how $B$ changes the size of the apparent shadow and how this effect is amplified by rotation.

			\subsection{Shadow Area}

			The shadow area is the area enclosed by the parametric curve $				\alpha =\alpha(r_p)$ and $\beta =\beta(r_p)$, where $r_p$ labels the unstable spherical photon orbits. In general, the shadow area is given by the contour integral
			\begin{equation}
				A_{sh}=\oint\beta d\alpha.
			\end{equation}
		For an equatorial observer with $\theta_o = \pi/2$, the shadow is symmetric about the $\alpha$-axis, and the contour integral reduces to the one-dimensional form
			\begin{equation}
			A_{\mathrm{sh}} =2\int_{r_p^-}^{r_p^+}\beta(r_p) \left|  \frac{d\alpha(r_p)}{dr_p} \right| dr_p,
			\end{equation}
			where $r_p^-$ and $r_p^+$ are the radii for which $\beta(r_p) = 0$, and the sign is chosen so that the integrand is positive. The factor of 2 is specific to the equatorial observer configuration and should not be used for general observer inclinations.
			
Since the metric functions depend on the Bertotti-Robinson parameter $B$ only through $B^2$, the shadow area admits the weak-field expansion
\begin{equation}\label{eq:shadow_expansion}
	A_{\mathrm{sh}}(B)
	=m^2\left[\mathcal{A}_0(j) + b^2\mathcal{A}_2(j) + \mathcal{O}(b^4)\right].
\end{equation}
where $\mathcal{A}_0(j)$ is the dimensionless Kerr shadow area and $\mathcal{A}_2(j)$ is the dimensionless leading-order magnetic correction.

To determine $A_2$, we expand the celestial coordinates as
\begin{equation}\label{eq:alphabeta}
	\begin{split}
		\alpha(r_p,B)
		&=
		\alpha_0(r_p)+B^2\alpha_2(r_p)+\mathcal O(B^4),
		\\
		\beta(r_p,B)
		&=
		\beta_0(r_p)+B^2\beta_2(r_p)+\mathcal O(B^4).
	\end{split}
\end{equation}

For an equatorial observer $\theta_0=\pi/2$ in the weak-field limit, the zeroth-order celestial coordinates simplify to $\alpha_0 = -\xi_0$ and $\beta_0 = \sqrt{\eta_0 - (a - \xi_0)^2}$. Using these relations, we have
\begin{equation}
	A_0^{\mathrm{dim}} = 2\int_{r_0^-}^{r_0^+} \beta_0(r_p)\frac{d\alpha_0}{dr_p}\,dr_p.
\end{equation}

Expanding about the Schwarzschild radius $r_p = 3m + \mathcal{O}(a)$ and performing the integration yields
\begin{equation}
	A_0^{\mathrm{dim}}=m^2\mathcal{A}_0(j) = 27\pi m^2 - \frac{3\pi}{2} a^2 + \mathcal{O}\left(\frac{a^4}{m^2}\right).
	\label{eq:A0}
\end{equation}
The leading term $27\pi m^2$ corresponds to the Schwarzschild shadow area, while the spin-dependent correction appears at order $j^2$.

Substituting the expansions from Eq.\eqref{eq:alphabeta} into the area integral gives the dimensionful $\mathcal{O}(B^2)$ coefficient
\begin{equation}
	A_2 ^{\mathrm{dim}}= 2\int_{r_0^-}^{r_0^+}\left[\beta_2(r_p)\frac{d\alpha_0}{dr_p} + \beta_0(r_p)\frac{d\alpha_2}{dr_p}\right]dr_p.
	\label{eq:A2_integral}
\end{equation}

The shifts of the integration limits $r_p^\pm$ do not contribute at this order because $\beta_0(r_0^-) = \beta_0(r_0^+) = 0$. Employing the relations
\begin{equation}
	\alpha_2 = -\xi_2,\qquad
	\beta_2 = \frac{\eta_2 + 2(a - \xi_0)\xi_2}{2\sqrt{\eta_0 - (a - \xi_0)^2}}.
\end{equation}
Restoring dimensions, the corresponding coefficient is
\begin{equation}
A_2^{\mathrm{dim}}=m^4\mathcal{A}_2(j) = -162\pi m^4- 192\pi a^2 m^2 + \mathcal{O}(a^4).
	\label{eq:A2_dimless}
\end{equation}

\begin{figure}[htbp]
	\centering
	
	\includegraphics[width=.750\textwidth]{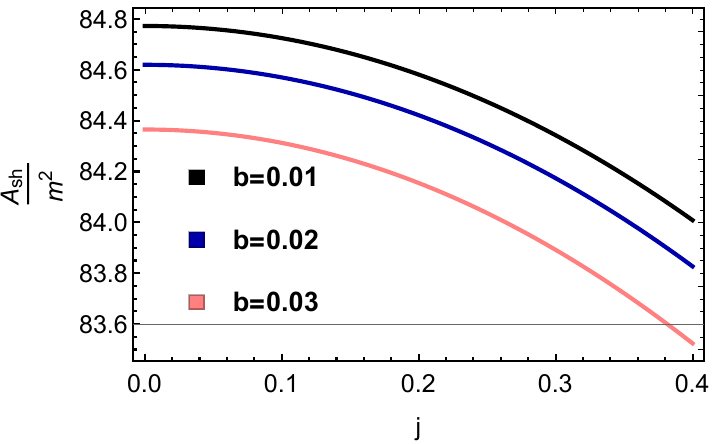}
	
	\caption{Dimensionless shadow area $A_{\rm sh}/m^2$ as a function
		of $j=a/m$, obtained from the weak-field and slow-rotation expansion
		in Eq.~\eqref{eq:final_area_dimless}. The plot is restricted to the perturbative range
		$j\ll1$ and $b\ll1$.}
	\label{fig:Ash}
\end{figure}

Substituting Eqs.~\eqref{eq:A0} and \eqref{eq:A2_dimless} into the expansion \eqref{eq:shadow_expansion}, we arrive at the final expression for the shadow area:
\begin{equation}
	A_{\mathrm{sh}}=A_0^{\mathrm{dim}}+B^2 A_2^{\mathrm{dim}}+\mathcal{O}(B^4),
\end{equation}
 in dimensionless form,
\begin{equation}
	\frac{A_{\mathrm{sh}}}{m^2} = 27\pi - \frac{3\pi}{2} j^2 - b^2\left(162\pi + 192\pi j^2\right) + \mathcal{O}(j^4, b^4).
	\label{eq:final_area_dimless}
\end{equation}
This result explicitly demonstrates that the Bertotti-Robinson magnetic background decreases the shadow area. 
Figure \eqref{fig:Ash} shows the dimensionless shadow area $A_{\rm sh}/m^2$ as a function of the spin parameter $j=a/m$ for different values of $b=Bm$. The shadow area decreases monotonically with increasing spin, reflecting the leading negative spin correction to the Kerr shadow. For fixed $j$, larger values of $b$ further reduce the shadow area. This confirms that the Bertotti-Robinson background produces a negative magnetic correction to the shadow size and that the reduction becomes stronger when the black hole is rotating.

\subsection{Magnetic Shadow Susceptibility}

The leading purely magnetic correction in Eq.~\eqref{eq:A2_dimless} is $\Delta A_{\mathrm{sh}}^{(B^2)} = -162\pi B^2 m^4,$ while the leading spin-magnetic cross term is
$\Delta A_{\mathrm{sh}}^{(B^2 a^2)} = -192\pi B^2 a^2 m^2.$

Thus, the reduction in shadow area becomes stronger for a nonzero rotation parameter $a$.

We define the magnetic susceptibility of the shadow area as
\begin{equation}
	\chi_B \equiv \frac{1}{A_{\mathrm{sh}}}\frac{\partial A_{\mathrm{sh}}}{\partial B}.
	\label{eq:chi_def}
\end{equation}

Using Eq.~\eqref{eq:final_area_dimless}, we find
\begin{equation}\label{eq:divAsh}
	\frac{\partial A_{\mathrm{sh}}}{\partial B} = -2B\left(162\pi m^4 + 192\pi a^2 m^2\right) + \mathcal{O}(B^3).
\end{equation}

Substituting Eq. \eqref{eq:final_area_dimless} into Eq.~\eqref{eq:chi_def}, and using  $j=a/m$ and $b=B m$, gives
\begin{equation}
	\frac{\chi_B}{m} = -\frac{2b\left(162 + 192j^2\right)}{27 - \frac{3}{2}j^2 - b^2\left(162 + 192j^2\right)}.
	\label{eq:chi_dimless}
\end{equation}
The negative sign of the magnetic shadow susceptibility, i.e., $\chi_{\rm B}<0$, has a clear physical interpretation: it quantitatively confirms that an increase in the external Bertotti-Robinson background parameter $B$ leads to a decrease in the black hole shadow area ($A_{\rm sh}$ decreases as $B$ increases). This behavior is consistent throughout the perturbative regime and is further amplified by the black hole's rotation.

\begin{figure}[htbp]
	\centering
	
	\includegraphics[width=.750\textwidth]{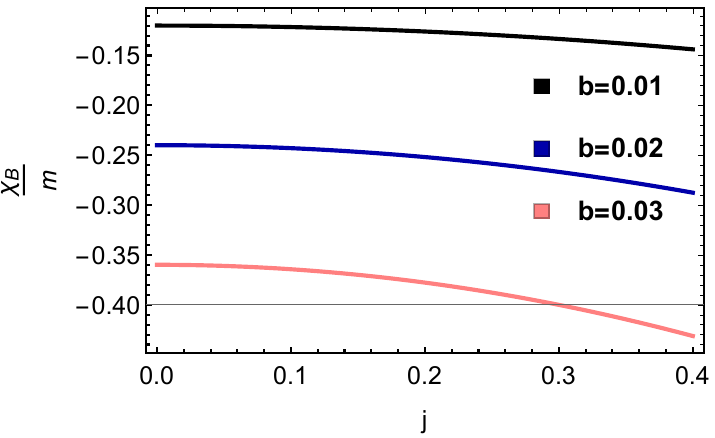}
	
	\caption{Dimensionless magnetic shadow susceptibility $ \chi_B/m $ as a function of $ j = a/m $ for different values of $ b = Bm $. The susceptibility is negative throughout the shown perturbative range. Indicating that increasing the Bertotti-Robinson background decreases the shadow area. Its magnitude increases with both $j$ and $b$, showing that rotation enhances the shadow's sensitivity to the external background.}
	\label{fig:Magneticshadowsuceptibikity}
\end{figure}

Figure \eqref{fig:Magneticshadowsuceptibikity} shows the dimensionless magnetic shadow susceptibility $\chi_B/m$ as a function of the spin parameter ($j=a/m$) for different values of $b=Bm$. The susceptibility is negative for all values of $j$, indicating that increasing the Bertotti-Robinson background parameter decreases the black hole shadow area. Moreover, $\chi_B/m$ becomes increasingly negative as the spin grows, indicating that rotation enhances the shadow's sensitivity to the external background. For fixed $j$, larger values of $b$ lead to a more negative susceptibility, which means that the shadow-area reduction becomes stronger in a stronger Bertotti-Robinson background. This behavior is consistent with the perturbative result that the leading magnetic correction to the shadow area is negative and is amplified by the spin-background coupling.

In the small-spin regime $j \ll 1$ with $b=Bm$ , this reduces to
\begin{equation}
	\frac{\chi_B}{m} \simeq -12b\left[1 + \frac{67}{54}j^2\right],
	\label{eq:chi_smallBj}
\end{equation}

This completes our perturbative calculation of the Kerr-Bertotti-Robinson shadow area in the weak-field and small-spin limit.

\section{Final Outcomes and Perspectives}
\label{sec:outlook}

In this work, we have presented a unified analysis of the thermodynamic and optical properties of Kerr--Bertotti--Robinson black holes. The parameter $B$ characterizes the external Bertotti--Robinson electromagnetic background and its gravitational backreaction. Our treatment encompasses the horizon structure, horizon thermodynamics, finite-surface Komar charges, null geodesics, spherical photon orbits, the ergosphere, and integrated shadow observables. In every sector, the standard Kerr results are recovered continuously in the limit $B\to0$.

Starting from the horizon condition $Q(r_h)=0$, which reduces to $\Delta(r_h)=0$, we obtained the outer horizon and inverted the resulting relation to express the effective horizon mass $M(r_h,a,B)$. In the weak-field regime, $Br_h\ll1$ and $Ba\ll1$, the leading deformation of the Kerr horizon relation appears at order $B^2$. Using the time normalization inherited from the Kerr-like coordinate $t$ and the regular-axis azimuthal normalization, we derived the Hawking temperature, horizon area, and entropy. Because the spacetime is not asymptotically flat, the horizon-defined mass should not be identified automatically with an ADM mass, and the geometrical angular velocity $\Omega_h$ should not be identified automatically with $(\partial M/\partial J)_{S,B}$. We therefore adopted a fixed-$a$ ensemble and introduced the Helmholtz-type free energy $F_a$ and the heat capacity $C_{a,B}$. The sign and divergences of $C_{a,B}$ provide local stability diagnostics, although a complete global phase structure would require a thermodynamic formulation based on independently derived conserved charges.

The occurrence of factors such as $1+B^2r^2$ also motivated a formal AdS-like parametrization of the external scale. In this description, $\mathcal{P}_B=3B^2/(8\pi)$ serves as an effective response variable, with associated fixed-$a$ quantities $V_B^{(a)}$ and $\Psi_B^{(a)}$. This construction is strictly a bookkeeping device: $B$ is not generated by a four-dimensional cosmological constant, the Bertotti--Robinson background is not asymptotically AdS$_4$, and these quantities are not the standard thermodynamic volume and magnetic conjugate of a fixed-$J$ first law. Establishing such a first law requires a separate construction of the physical mass and angular momentum.

The extremal configuration follows from $T_h=0$. We found $r_{\rm rem}=a/\sqrt{1-a^2B^2}$ and $m_{\rm rem}=a(1-\frac{1}{2}B^2a^2)/\sqrt{1-B^2a^2}$. Thus, the extremal radius receives its leading correction at order $B^2$, whereas the remnant mass is unchanged at that order and first departs from its Kerr value at order $B^4$. Regularity of the physical branch requires $P_0>0$ together with $|Ba|<1$; the latter condition prevents the extremal radius and mass from becoming singular.

The Komar analysis further illustrates the subtleties associated with gravitational charges in this non-asymptotically flat geometry. The finite-radius Komar integral associated with the stationary Killing vector reduces to $m$ in the Kerr limit, but for $B\neq0$ it acquires explicit $B^2$ corrections and depends on the integration radius. This surface dependence reflects the contribution of the external electromagnetic background and its backreaction between different integration surfaces. Consequently, the finite-surface Komar quantity cannot be interpreted as a unique ADM-like black-hole mass without an appropriate charge prescription and background subtraction. By contrast, the Komar charge of the horizon generator, $K_\chi=\kappa_hA_h/(4\pi)=2T_hS_h$, represents a horizon quantity rather than the total stationary mass; this distinction remains present even in the rotating Kerr limit.

In the optical sector, the Hamilton--Jacobi equation separates for null geodesics, allowing us to derive the radial and angular potentials, the impact parameters of spherical photon orbits, and the photon-region boundaries. We also obtained celestial coordinates for a static observer at finite radius. The simplified observer expressions apply in the intermediate regime $m\ll r_o\ll B^{-1}$ and should not be interpreted as a conventional asymptotic-observer limit, because the spacetime is neither asymptotically flat nor asymptotically AdS.

Within the weak-field expansion, supplemented by a slow-rotation expansion where required, the Bertotti--Robinson background produces several consistent optical trends. In the parameter range examined, it shifts the photon-region boundaries outward, decreases the angularly averaged ergosphere thickness, increases the photon--ergosphere gap, and reduces the shadow area. Rotation generates the expected prograde--retrograde splitting and introduces mixed spin--background corrections that enhance this asymmetry. For $B$ treated as a nonnegative background strength, the magnetic shadow susceptibility is negative, confirming that the shadow area decreases as $B$ increases; its magnitude is enhanced by rotation. These conclusions are perturbative and should not be extrapolated without further analysis to strong backgrounds or near-extremal spins.

Taken together, these results show that the Bertotti--Robinson background affects not only local horizon thermodynamics but also the relation among the horizon, ergosphere, photon region, and observed shadow. The quantities introduced here---in particular, the fixed-$a$ response coefficients, the averaged ergosphere thickness, the photon--ergosphere gap, the shadow area, and the magnetic shadow susceptibility---provide a systematic set of diagnostics for comparing Kerr--Bertotti--Robinson black holes with their Kerr counterparts.

Several extensions follow naturally. A primary objective is to construct the conserved mass and angular momentum using a covariant phase-space method, including the necessary background subtraction and integrability analysis. This would make it possible to test a fixed-$J$ first law and to determine how the fixed-$a$ response quantities introduced here are related to genuine thermodynamic conjugates. A complementary nonperturbative study should map the extremal and stability structure over the full admissible parameter space, including stronger backgrounds and near-extremal rotation. Full numerical ray tracing for finite-distance observers, generic inclinations, accretion disks, and realistic emission models will be required to determine how the predicted changes in shadow area and photon--ergosphere separation appear in synthetic images. Finally, connecting the dimensionless parameters to astrophysical field strengths, quantifying degeneracies with spin, inclination, and plasma effects, and exploring possible links with effective-medium and negative-refraction descriptions of spacetime optics constitute important directions for future work \cite{Orlando}.
\section*{Acknowledgments}
Funding comes partly from the FY2024-SGP-1-STMM Faculty Development Competitive Research Grant (FDCRGP) no.201223FD8824 and SSH20224004 at Nazarbayev University in Qazaqstan. OL acknowledges financial support from the National Institute for Astrophysics (INAF) of Brera.

	\end{document}